\documentclass[aps,pre,tightenlines,nofootinbib,showpacs,superscriptaddress,longbibliography,twocolumn]{revtex4-1}

\usepackage{amsmath}
\usepackage{amssymb,amsfonts,latexsym}
\usepackage{graphicx}
\usepackage[english]{babel}
\usepackage{color}
\usepackage{textcomp}
\usepackage{multirow}
\usepackage{lineno}
\usepackage{subfigure}

\def \bea{\begin{eqnarray}}
\def \eea{\end{eqnarray}}
\def \non{\nonumber}

\setcitestyle{super}

\newcommand{\mvh}[1]{\textcolor{black}{#1}}

\begin{document}

\title{Combinatorial Design of Textured Mechanical Metamaterials}

\author{Corentin Coulais}
\affiliation{Huygens-Kamerlingh Onnes Lab, Universiteit Leiden, PObox 9504, 2300 RA Leiden, The Netherlands}
\affiliation{FOM Institute AMOLF, Science Park 104, 1098 XG Amsterdam, The Netherlands}
\author{Eial Teomy}
\affiliation{School of Mechanical Engineering and The Sackler Center for Computational Molecular and Materials Science, Tel Aviv University, Tel Aviv 69978, Israel}
\author{Koen de Reus}
\affiliation{Huygens-Kamerlingh Onnes Lab, Universiteit Leiden, PObox 9504, 2300 RA Leiden, The Netherlands}
\author{Yair Shokef}
\affiliation{School of Mechanical Engineering and The Sackler Center for Computational Molecular and Materials Science, Tel Aviv University, Tel Aviv 69978, Israel}
\author{Martin van Hecke}
\affiliation{Huygens-Kamerlingh Onnes Lab, Universiteit Leiden, PObox 9504, 2300 RA Leiden, The Netherlands}
\affiliation{FOM Institute AMOLF, Science Park 104, 1098 XG Amsterdam, The Netherlands}

\maketitle

{\bf The structural complexity of metamaterials is limitless, although in practice, most designs comprise periodic architectures which lead to materials with spatially homogeneous features \cite{Lakes_science1987,mullinPRL,Grima,Schaedler_science2011,Nicolaou_NATMAT2012,Guest_pnas2013,Babaee_AdvM2013,Wegener_reviewRPP2008,Florijn_PRL2014,Coulais_PRL2015,Waitukaitis_PRL2015}.
More advanced tasks,  arising in e.g. soft robotics, prosthetics and wearable tech, involve spatially textured \mvh{mechanical} functionality
which require aperiodic architectures. However, a na\"ive  implementation of such structural complexity invariably leads to frustration, which prevents coherent operation and impedes functionality. Here we introduce a combinatorial strategy for the design of aperiodic yet frustration-free \mvh{mechanical} metamaterials, whom we show to exhibit spatially textured functionalities. We implement this strategy using cubic building blocks - voxels - which deform anisotropically, a local stacking rule which allows cooperative shape changes by guaranteeing
that deformed building blocks fit as in a 3D jigsaw puzzle, and 3D printing. We show that, first, these aperiodic metamaterials exhibit long-range holographic order, where the 2D pixelated surface texture dictates the 3D interior voxel arrangement. Second, they act as programmable shape shifters, morphing into spatially complex but predictable and de\-sign\-able shapes when uniaxially
compressed. Third, their mechanical response to compression by a textured surface reveals their ability to perform
sensing and pattern analysis. Combinatorial design thus opens a new avenue towards \mvh{mechanical} metamaterials with unusual order and machine-like
functionalities.
}

The architecture of a material is crucial for its properties and functionality. This connection between form and function is leveraged by mechanical metamaterials\cite{Lakes_science1987,mullinPRL,Grima,Schaedler_science2011,Nicolaou_NATMAT2012,Guest_pnas2013,Babaee_AdvM2013,Wegener_reviewRPP2008,Florijn_PRL2014,Coulais_PRL2015,Waitukaitis_PRL2015,Liu_Science2000,Buckmann_Natcomm2014,Silverberg_science2014,Paulose_PNAS2015},
whose patterned microstructures are designed to obtain unusual behaviors such as negative response parameters\cite{Lakes_science1987,Coulais_PRL2015}, multistability\cite{Nicolaou_NATMAT2012,Florijn_PRL2014,Waitukaitis_PRL2015} or programmability\cite{Silverberg_science2014,Florijn_PRL2014}. For ordinary
materials,  aperiodic architectures and structural complexity are associated with \mvh{geometric} frustration \mvh{(local constraints cannot be satisfied everywhere \cite{Wannier,FrustrationBook})} which prevents a coherent and predictable response.
Frustration hinders functionality, and  metamaterial designs have thus focused on periodic structures. However, the exquisite control provided by 3D printing provokes the question whether one can design and create structurally complex yet frustration-free metamaterials\cite{lipson_robotevolution}.

\begin{figure}[b!]
\centerline{\includegraphics[width=0.99\columnwidth,clip,trim=1.8cm 5.2cm 11.5cm 2.cm]{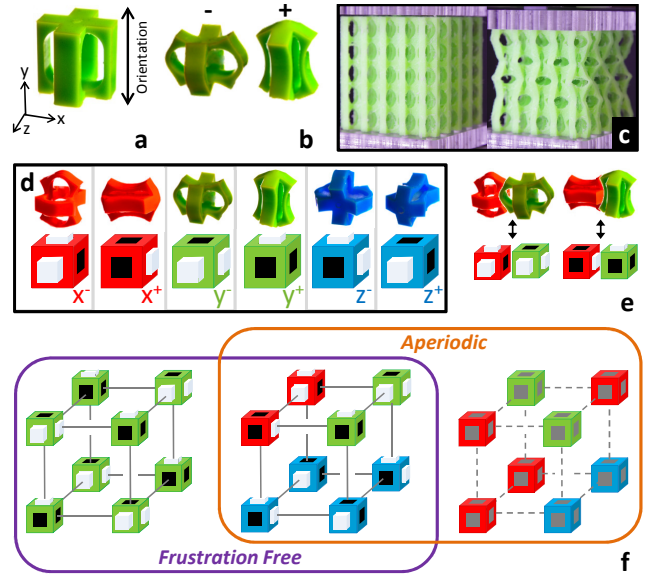}}
\caption{\mvh{Voxelated Mechanical Metamaterials} (a) Flexible anisotropic building block in its undeformed state. (b)
Corresponding flattened (-) and elongated (+)
deformed bricks. (c)
A $5\times5\times5$ metacube consisting of parallel blocks shows
a collective deformation under uniaxial compression.
(d) Bricks and their schematic representation, where
colour indicates orientation, and black dents and white protrusions represent deformations. (e) Adjacent bricks fit when appropriately polarized.
(f) Periodic, complex and frustrated $2\times 2\times 2$ stackings (from left to right) - for the
latter no consistent brick configuration exist (grey/dashed). Schematic symbols are separated from one another for visualization purposes.
}
\end{figure}

We foray into this unexplored territory using a combinatorial design strategy. We assemble 1 cm$^3$ flexible, cubic
{ \em building blocks} or voxels into a cubic lattice which then forms a metamaterial (Fig.~1a). These  building blocks are anisotropic and  have one soft mode of deformation aligned with its internal axis of orientation, resulting in elongated or flattened shapes that we refer to as {\em bricks} with positive or negative polarization (Fig.~1b). Generally, \mvh{mechanical} metamaterials with randomly orientated building blocks are frustrated, as it is impossible for all blocks to cooperatively deform according to their soft mode: the bricks do not fit.
We call voxelated\cite{Ware_science2015} metamaterials that allow soft deformations, or equivalently,
where all bricks fit, {\em compatible}. A trivial example of a compatible configuration is a periodic stacking of alternatively polarized, parallely orientated bricks. Hence, a periodic metamaterial consisting of parallel blocks is expected to exhibit a collective and harmonious deformation mode. We realized such \mvh{a} metamaterial by a combination of 3D printing and moulding (see Methods). Uniaxial compression indeed triggers a collective pattern change\cite{mullinPRL} in three dimensions, and produces the expected staggered configuration where each brick is adjacent to six bricks of opposite polarization (Fig.~1c\mvh{; see video 1 described in the S.I.}).

We now consider how to design aperiodic yet frustration-free \mvh{mechanical} metamaterials. Crucially, the internal structure of our blocks is anisotropic, and each block can be oriented independently to allow structurally complex architectures. We think of these blocks as voxels, represent their orientation at each lattice \mvh{point} with a colour (Fig.~1d), and explore the discrete design space of such voxelated metamaterials. Compatibility requires two conditions to be met. First, pairs of neighbouring bricks should exhibit closely matching shapes along their common face. Our building blocks are precisely designed such that given the polarization of one brick, the polarization of an adjacent brick can be adapted so that the pair have a tight fit - irrespective of their mutual orientation. Hence, we only need to track the outward or inward deformations of the surfaces of the building blocks (Fig.~1e). The second compatibility condition concerns the
combinatorics of the voxel arrangement: all bricks should fit, such that protrusions and depressions of all neighbouring bricks are matched. In general, the first condition can be met by clever building block design, while the second condition leads to a thorny combinatorial 3D tiling problem.

\begin{figure*}[t!]
\centerline{\includegraphics[width=1.6\columnwidth ,clip,trim=1.8cm 1cm 4.5cm 1.45cm]{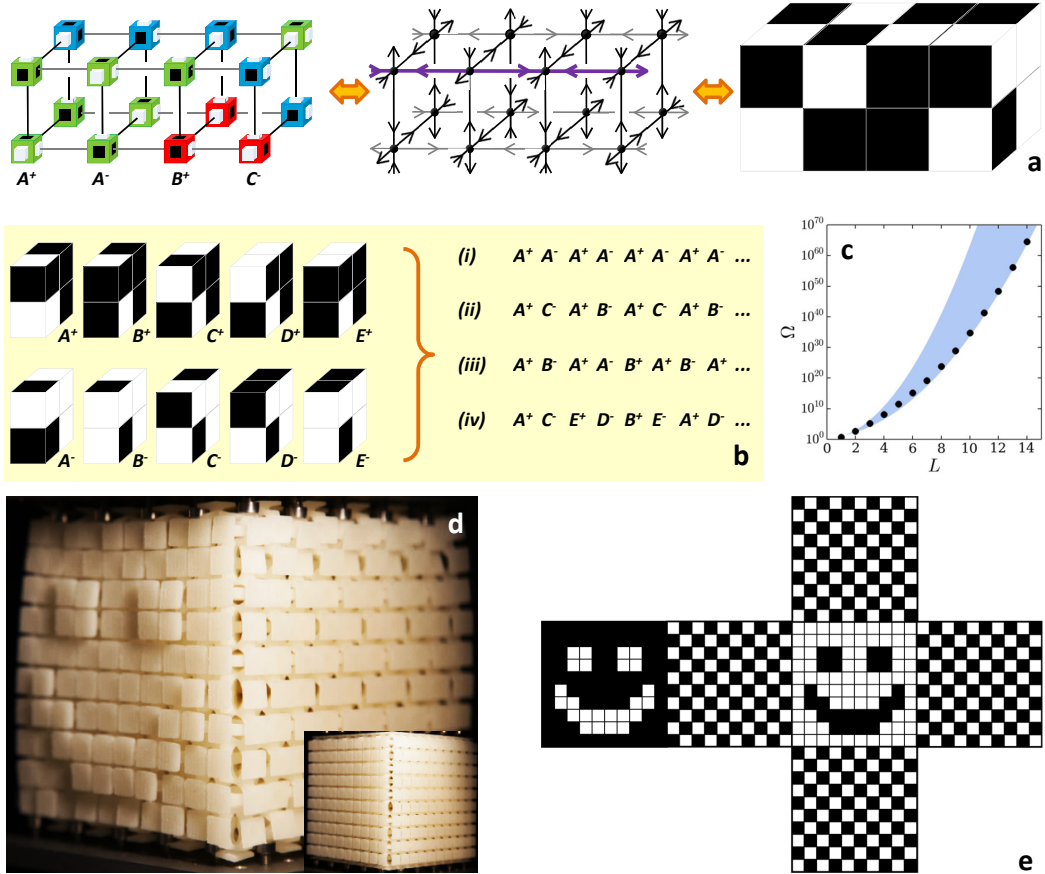}}
\caption{\mvh{Combinatorial Design} (a) Mapping of bricks to internal spins to surface spins (from left to right). (b) Dictionary of five pairs of motifs
compatible with a given $x$-texture, and examples of (i-ii) periodic, (iii) quasiperiodic (beginning of Fibonacci sequence) and
(iv) aperiodic motif stackings. (c) Exact number of compatible $L\times L\times L$ spin configurations $\Omega$ (dots) and
lower and upper bounds (blue region) - see Methods and S.I.
(d) A $10\times 10 \times 10$ metacube  reveals
its precisely designed surface texture under uniaxial compression.
Square surface pedestals added for visualization. Inset: undeformed metacube.
See Methods for experimental details and 
\mvh{videos 2-4 described in the S.I.}. (e) Schematic representation of the deformations at all surfaces of this
metacube.}
\label{fig:fig2}
\end{figure*}

As we will show, while the compatibility condition is violated in most random configurations \mvh{which are thus frustrated},
our specific building blocks allow for a plethora of complex configurations where all protrusions and depressions match.
These non-parallel, structurally complex yet compatible architectures compound the rich spatial texture of aperiodic materials with
the predictability of ordered materials and form the blueprint for aperiodic, frustration free \mvh{mechanical} metamaterials (Fig.~1f).

The design of complex architectures is simplified by mapping brick configurations to spin-configurations which satisfy a so-called ice-rule\cite{Harris,Wang_Nature2006,Castelnovo_Nature2008,Nisoli_RMP2013}, and as such is reminiscent of
tiling~\cite{tilingbook} and constraint satisfaction~\cite{Kirkpatrick_Science1983,Mezard_book} problems.
We identify each brick with a vertex, connected to neighbouring vertices by bonds which represent the common face between bricks (Fig.~2a). Dents and bumps map to inward or outward spins $\sigma_x, \sigma_y, \sigma_z $, and by placing a single spin per bond, the first compatibility condition is trivially satisfied. The second condition maps to the ice-rule, which stipulates that the six bonds of each vertex should correspond to a brick configuration, where the six bricks $x^-,\dots z^+$
correspond to spin configurations $(\sigma_x, \sigma_y, \sigma_z) = (-++),(+--),(+-+),(-+-),(++-),(--+)$.
Evidently, each allowed spin configuration corresponds to a compatible brick stacking and corresponding voxel configuration.
We note that, conversely, each compatible voxel configuration corresponds to two spin configurations related by {\em parity} (spin flip), a symmetry which originates from the opposite polarizations allowed by each building block.

All compatible metamaterials thus obtained feature an unusual form of long range order which relates the boundary to the bulk.
Due to the bricks' reflection symmetry, spins along lines of bonds are alternating. Therefore, spins at opposing boundaries are equal
(opposite) when their distance is odd (even). As spins at the surface of a metamaterial represent its texture of bumps and dents,
this implies that textures at opposite faces of a metacube are directly linked. Moreover, once the surface texture is fixed, all internal spins and therefore bricks are determined (Fig.~2a). We call this unusual relation between surface and bulk ``holographic order'' \mvh{(see Methods)}.

The combination of parity and holographic order implies that any compatible $n\times p\times q$ {\em motif} can be
stacked in a space filling manner\mvh{, as the surface spins of such motifs have compatible textures}.
Moreover, once the $x$-spins are fixed along a plane, we can determine
a dictionary containing all motifs $A^+, B^+,\dots$ with matching $x$-spins, and by parity obtain $A^-,B^-,\dots$
(Fig.~2b \mvh{and Methods}). These can be stacked in arbitrary order, as long as we alternate between '+' and '-' motifs; this allows the
straightforward design of periodic, quasiperiodic and aperiodic metamaterials (Fig.~2b). \mvh{By removing building blocks at the
boundary, complex shapes can be realized, but for simplicity we focus here on cubic metamaterials.}

Holographic order significantly restricts the number of potential compatible configurations: while for
generic configurations their multitude is set by the volume, for compatible configurations it is set by the surface area.
Moreover, many surface textures lead to forbidden internal vertices, e.g. where all spins are equal. \mvh{For example, in general it is not possible to arbitrarily choose the surface texture at two faces simultaneously.}  Nevertheless, the number of
distinct $L\times L \times L$ spin configurations $\Omega(L)$ is astronomical. To quantify the design limits and possibilities, we exactly evaluated $\Omega(L)$ up to $L=14$ where $\Omega \approx 3\times10^{64}$, and obtained strict and asymptotic lower and upper bounds (Fig.~2c, Methods and S.I.).

Despite the limitations imposed by compatibility, the design space of voxelated metamaterials is huge.
To illustrate this, we have constructed a general algorithm to obtain all $L\times L \times 1$ motifs compatible with a given texture $\{\sigma_z\}$  (see Methods and S.I.); for each texture there are at least two distinct motifs. We show now that we can use this to design arbitrarily pixelated patterns of bumps and depressions, or textures, at a given surface of a metacube as a step towards arbitrary shape morphing materials. In Fig.~2d we show a rationally designed metacube  created by 3D printing. Under uniaxial compression (see Methods), the initially flat surface of this cube reveals its spatial texture, with the front and back related by holography (Fig.~2e and Methods\mvh{; See videos 2-4, described in the S.I.}). This cooperative, complex yet controlled shape morphing
illustrates that our combinatorial method allows for the rational design of shape shifting metamaterials.

\begin{figure*}[t!]
\centerline{\includegraphics[width=1.9\columnwidth,clip,trim=1.0cm 1.2cm 1.2cm 0.4cm]{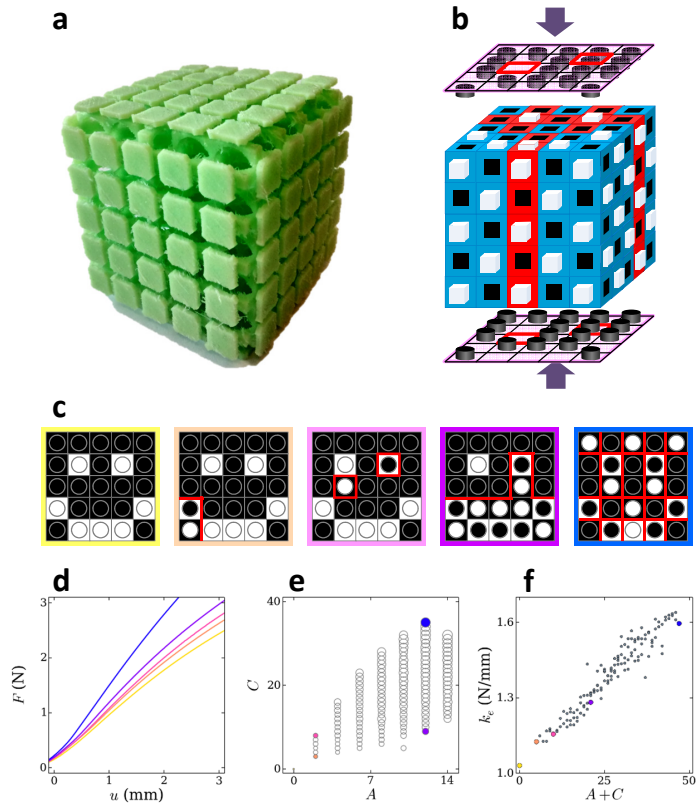}}
\caption{ Pattern recognition and pattern analysis. (a) Experimental realization
of an elastic $5\times 5\times5$ metacube programmed for a smiley texture.
(b) Schematic
of experiments where this metacube  is
compressed between patterned clamps. (c)
Examples of mismatch between cube lock texture $\sigma_L$ (squares) and boundary key textures $\sigma_k$ (circles). The boundary between regions of opposite parity is indicated in red. (d)
Experimental force-compression curves. Colour corresponds to key textures in panel (c).
(e) The experimentally obtained stiffness $k_e$ varies systematically with both the area $(A)$ and circumference $(C)$ of the mismatch. Coloured data points  correspond to key textures in panel (c), and size of the circle represents $k_e$.
(f) The stiffness $k_e$ is essentially linear in
  $A+C$. }
\label{fig:fig3}
\end{figure*}

We finally show that when aperiodic metacubes are compressed by patterned surfaces their response can be employed for mechanical pattern analysis. We created a compatible  $5\times 5 \times 5$ metacube, programmed with a smiley texture  $\pm \{\sigma_z^L\}$
which acts as a ``lock'' (see Methods), which is compressed between two identical surfaces that have a pixelated ``key'' texture $\sigma_z^K$ created by placing eighteen stubs  in templated clamps (Fig.~3a-b). We characterize the difference between lock and key patterns by  the  area or number of misplaced stubs, $A$, as well as the circumference of the misplaced area, $C$ (Fig.~3c), and use 136 different key patterns which cover all possible values of $A$ and $C$ \mvh{that can be reached by 18 stubs}. For each key, we performed experiments (respectively simulations) to determine the stiffness $k_e$ (respectively $k_s$) via the slope of the force-compression curves - both values agree very closely (see Methods). When the key equals one of the two lock textures $\pm \{\sigma_z^L\}$, all bricks deform compatibly and $k$ is low. Incompatibly textured surfaces push metacubes into frustrated states, leading to an energy penalty and increased stiffness (Fig.~3d).
The increase with $A$ evidences simple lock and key functionality, but when $k$ is plotted as
function of $A$ and $C$, the stiffness is seen to increase with $C$ also - for the same number of misplaced stubs,
a range of stiffnesses can be observed (Fig.~3e). When plotted as function of $A+C$, all our data collapses on a straight line,
which evidences intricate collective phenomena at play (Fig.~3f). We suggest that due to parity, different parts of the cube deform
in opposing parity, and that the stiffness is determined by the size of the domain walls separating these regions, which is given by $A+C$ (see methods). Together, this demonstrates the ability of a metacube to perform an arithmetical calculation on the mismatch between key and lock patterns, in behaviour more readily associated with machines than with materials.

\mvh{Combinatorial strategies open up the design of machine materials which can be programmed
with specific shape sensing and shape shifting tasks. We anticipate that combinatorial design of textured metamaterials can be extended in various directions. First, the
inclusion of vacancies could lead to multishape materials\cite{Cho_PNAS2014}, whereas defects can induce controlled frustration to obtain multistability, memory and programmability\cite{Florijn_PRL2014,Silverberg_science2014,Waitukaitis_PRL2015}.
Second, differently shaped building blocks such as triangles or hexagons in two dimensions,
truncated octahedra and gyrobifastigii in three dimensions, or mixtures of building blocks could be used
to tile space. Third, building blocks with degrees of freedom different from the simple 'inwards or outwards' deformations
considered here could be considered --- a prime example being  origami units that have folding motions ~\cite{Guest_pnas2013,Silverberg_science2014,Waitukaitis_PRL2015}.
Finally, heating or magnetic fields instead of compression could be used
to actuate shape changing metamaterials,
while non-mechanical textured functionalities such as wavefront shaping could also be achieved.
We envision a range of applications where control and processing of spatially complex mechanical information
is key.  Textured metamaterials can be designed to naturally interface with the
complex shapes and shapeability of the human body, in prosthetics, haptic devices, and wearables. Moreover,
shape changing is central to a wide variety of actuators and sensors, in particular in the context of soft robots
\cite{Gracias_PNAS2009,Whitesides_PNAS2011,Overvelde_PNAS2015}. Finally, at smaller scales, controllable surface textures could control friction, wetting, and drag.}


\def\bibsection{\section{\refname}}


\section{Acknowledgements}
We are grateful to J. Mesman for technical support. We like to thank \mvh{R. Golkov, Y. Kamir, G. Kosa, K. Kuipers, 
F. Leoni, W. Noorduin and  V.Vitelli} for discussions.
We acknowledge funding from the Netherlands Organisation for Scientific Research grants VICI No NWO-680-47-609 (M.v.H. and C.C.)
and VENI No NWO-680-47-445 (C.C.) and the Israel Science Foundation grants No. 617/12, 1730/12 (E.T and Y.S.).

\section{Author Contributions}
C.C. and M.v.H. conceived the main concepts. C.C., E.T., Y.S. and M.v.H. formulated the spin-problem. E.T. and Y.S. solved the spin-problem.
C.C and K.d.R. performed the experiments and simulations with inputs from E.T., Y.S. and M.v.H. C.C. and M.v.H wrote the manuscript with contributions from all authors.

\section{Author Information}
Reprints and permissions information is available online at www.nature.com/reprints.
The authors declare no competing financial interests.
Correspondence and requests for materials should be addressed to C.C. (coulais@amolf.nl).

\section{Methods}

\textbf{Combinatorial Design.}
The presence of holographic order reduces the number of potential compatible $L\times L\times L$ spin configurations from $2\times 3^{L^3}$ ($L^3$ blocks
with 3 orientations and parity) to $2^{3L^2}$ (3 pairs of opposing surfaces with each $L^2$ spins). The vast majority of these are not compatible and to better understand our design space,
we discuss how to exactly evaluate and obtain upper and lower bounds for $\Omega(L)$,
the number of $L$-cubes (short for the number of potentially compatible $L\times L\times L$ spin configurations). We construct cubes by stacking motifs, and $Q$
counts the number of $L$-motifs (short for $L\times L \times 1$ motifs) compatible with a given texture $\left\{\sigma_{z}\right\}$.

To understand the possible motifs,
we now classify the patterns of $z$-bricks that arise in $L$-motifs.
Crucially, $z$-bricks are sources or sinks for the in-plane spins $\sigma_x$ and $\sigma_y$, and therefore each $2\times 2$ submotif can only contain 0, 2 or 4  $z$-bricks.
This restricts the patterns of $z$ bricks to columns, rows, and intersecting columns and rows.
In general, we can enumerate the patterns of $z$-bricks using
binary vectors $c_i$ and $r_j$, and placing a $z$-brick at location $(i,j)$ only when $c_i \ne r_j$ (\mvh{Fig.~\ref{fig:combi}a}).
On the $z$-bricks, the $z$-spins form checkerboard patterns, whereas the spins in the remainder of the pattern can be chosen at will by filling each position with either an $x$ or $y$ brick.

In the absence of $z$-bricks, we can obtain two motifs by
fixing $\sigma_x$ and $\sigma_y$  to be opposite and
alternating, i.e., at site $(i,j)$ of the motif
$\sigma_x=-\sigma_y=(-1)^{i+j}$ or $\sigma_x=-\sigma_y=(-1)^{i+j+1}$. In both cases,
each vertex is compatible with either an $x$ or a $y$-brick, with corresponding positive or negative $\sigma_z$. This allows the straightforward design of two $L\times L \times 1$ motifs consistent with any $\{\sigma_z\}$.
In \mvh{Fig.~\ref{fig:combi}}b we show these motifs, as well as four more  that are compatible
with a $5\times 5$ smiley texture - hence $Q=6$ for this texture.
In principle, $\Omega(L)$ can be exactly evaluated by determining for each texture $\{\sigma_z\}$ the number $Q$, and then summing over all textures (see S.I. and Fig.~\mvh{\ref{fig:combi}}c):
\begin{equation}\label{naivecount}
\Omega(L) := \Sigma_{\{\sigma_z\} } Q^L~.
\end{equation}

{\bf Lower Bound.} A lower bound for $\Omega$ follows from
our construction to create two motifs for any spin configuration, which implies that $Q\ge 2$. As these can be stacked in arbitrary order, this yields at least $2^{L}$ spin-configurations for a given texture.
Since there are $2^{L^{2}}$ ${\sigma}_{z}$ textures, we find that
\begin{equation}
\Omega(L)\geq  2^{L^{2}+L}~.
\end{equation}

{\bf Staggered Spins.}
To simplify the counting of the number of compatible spin-configurations $\Omega(L)$ (for details see S.I.), we
define staggered spins $\tilde{\sigma}$, such that for site $(i,j,k)$ in the
metacube $\tilde{\sigma}^{(i,j,k)}_{\alpha}=\left(-1\right)^{i+j+k}\sigma^{(i,j,k)}_{\alpha}$, for $\alpha=x,y,z$. Under this invertible
transformation, a checkerboard texture of $\left\{\sigma_{z}\right\}$, for example, corresponds to a homogeneous texture of $\left\{\tilde{\sigma}_{z}\right\}$ where
all staggered spins are equal to either $+1$ or $-1$. Moreover, all sites in a given row, column or tube have the same value of the $\tilde{\sigma}_{x}$,  $\tilde{\sigma}_{y}$ or $\tilde{\sigma}_{z}$, respectively.

{\bf Upper Bound.} For a simple upper bound we note that the maximum value of $Q$ is obtained when
$\tilde{\sigma}_{z}\equiv+1$ or $\tilde{\sigma}_{z}\equiv-1$. For each of
these textures there are $Q=2^{L+1}-1$ spin configurations. Consider for example $\tilde{\sigma}_{z}\equiv+1$.
If $\tilde{\sigma}_{x}\equiv-1$ then all the $\tilde{\sigma}_{y}$ are free, leading to $2^{L}$ compatible structures,
if $\tilde{\sigma}_{y}\equiv-1$ then all the $\tilde{\sigma}_{x}$ are free, leading to an additional $2^{L}$ compatible structures.
As $\tilde{\sigma}_{x}\equiv\tilde{\sigma}_{y}\equiv-1$ was counted twice, the total number is $2^L+2^L-1=2^{L+1}-1$. Hence, we obtain as upper bound:
\begin{equation}
\Omega(L) \le 2^{L^2} \times (2^{L+1}-1)^L < 2^{2L^2+L}~.
\end{equation}
A stricter upper bound is derived in the S.I.: $\Omega(L) \leq 4L\cdot (3/4)^L\cdot 2^{2L^2} $. The exact evaluation of $\Omega(L)$ is explained in the S.I. and the results are given in Table~\mvh{EDT1}.

{\bf Design Limits.}
We note here that if $n$ is the number of pairs of opposing surfaces where the
spins can be chosen freely, $\Omega(L) \approx 2^{n L^2}$, and that the simple
lower and upper bound given above roughly correspond to $n=1$ and $n=2$.
Approximate calculations detailed in the S.I. lead to $2^{L^2+L+log_2(3)} \leq
\Omega \leq 2^{L^2+2L+2}$, and our exact evaluation of $\Omega(L)$ shows that
for large $L$, $\Omega$ is quite close to the lower bound (Fig.~2a). Hence, once
the texture of one surface is fixed, there is limited freedom, apart from the
stacking order of motifs, to design textures at other surfaces. For most spin
textures, only the two simple motifs are compatible, and $z$-bricks play a minor
role.

\textbf{Materials and Fabrication.}
We created the $5\times5\times5$ specimens by 3D printing water-soluble moulds, in which we cast a well
calibrated silicon rubber (PolyvinylSiloxane, Elite Double 32, Zhermarck, Young's Modulus, $E=1.32$~MPa, Poisson's ratio $\nu\sim0.5$).
The unit bricks measure $11.46 ~ \textrm{mm} \times 11.46 ~\textrm{mm}\times 11.46~ \textrm{mm}$, with a spherical
pore of diameter $D=10.92$~mm in the center and four cuboid inclusions of dimension $4.20\textrm{~mm}\times4.20\textrm{~mm}\times11.46\textrm{~mm}$
at the $x$ and $y$ corners --- See Fig.~\mvh{\ref{fig:geometry}}a. They are stacked with a pitch of $a=11.46$~mm, such that the filaments between the unit cells
have a non homogeneous cross-section with a minimal width $d=0.54$~mm and a depth $w=3.6$~mm --- See Fig.~\ref{fig:geometry}b.

The $10\times10\times10$ sample has the same brick dimensions and was 3D printed commercially (Materialise, Leuven, Belgium) out of sintered
polyurethane ($E\approx 14$~MPa). On the faces of the aperiodic cubes, square pedestals were added to facilitate both visualization of the surface texture and compression by textured boundaries.

\textbf{Mechanical tests.}
We compressed both metacubes
at a rate of $0.02$~mm/s  in a uniaxial testing device (Instron type 5965)
which controls the compressive displacement $u$ better than $10\,\mu$m  and measures
the compressive force with a $10^{-4}$~N accuracy at an acquisition rate of $0.5$~Hz.

While we used flat plates for Fig.~1c, we used textured boundary conditions for Figs.~2d and 3.
We created textured top and bottom boundaries using aluminium plates with female $3$~mm threads positioned in a square array of pitch $p=11.46\pm0.02$~mm, in which we mount
stainless steel $M3$ screws whose caps were machined to a height of $3.50\pm0.01$~mm \mvh{(See \mvh{Fig.\ref{fig:clamp}}a)} - this ensures
precise levelling of the pins and flexibility in texture. At the start of each experiment,
the cubes were gingerly positioned and aligned by hand within a $1$~mm accuracy on the bottom boundary.
The screws were placed to form identical (respectively complementary) top and bottom patterns
for the $5\times5\times5$ (respectively $10\times10\times10$) cube.

For the $10\times10\times10$ cube, designed as in Fig.~\mvh{\ref{fig:antismiley}}a, we used checkerboard textures
leading to the desired pattern on one face (Fig.~2), the reverted pattern on
the opposite face (Fig.~2 and ~\mvh{\ref{fig:antismiley}}b) and checkerboard textures on the other faces (see Fig.~2 and \mvh{\ref{fig:antismiley}}c).

\textbf{Numerical Simulations.}
We probed the response of a $5\times5\times 5$  aperiodic smiley metacube to different textures by performing a fully nonlinear analysis within the
commercial package \textsc{Abaqus/Standard}. We modelled the elastomer using a neo-Hookean strain energy density
with a Young's modulus $E=1.32$~MPa and a Poisson's ratio $\nu=0.4999$. We carried out a mesh optimization and a mesh density study leading to a typical mesh size of
$0.6$~mm and a total number of $1.5\times10^6$ hybrid quadratic tetrahedral elements (Abaqus type C3D10H). We applied uniaxial compression by applying
10  steps of magnitude $\Delta u=0.25$~mm, using the exact same boundary conditions and dimensions as in the experiments (Fig.~\mvh{\ref{fig:barcodedetails}}b).

\textbf{Determination of $k$.}
The numerical force-displacement curves are very well fitted by the quadratic form $F(u)=k u +\eta u^2$, which captures the effect of the nonlinear softening and
which allows
an accurate estimation of the stiffness $k$.
The experimental determination of the stiffness required special care, as  small gravity-induced sagging of the cube causes a soft knee in the force displacement curve
when the top boundary makes
contact with the sample. Therefore, we determined the stiffness by fitting  the force-displacement
curves to the same quadratic function as for the numerics, focussing on intermediate displacements
$0.8$~mm $\le u \le 2.5$~mm away from
the knee  where $dF/du$ is linear in $u$  --- see Fig.~\mvh{\ref{fig:barcodedetails}}a.

\textbf{Lock and Key Mechanics.}
For the lock and key experiments, we used a
$5\times5\times5$ cube made by stacking 5 $B^{\pm}$configurations (\mvh{Fig.~\ref{fig:combi}}b). The key patterns consisted
of $18$ screws placed in a $5\times 5$ array (see Fig.~3b-c \mvh{and \ref{fig:clamp}a-b}). We focused on $136$ configurations that have distinct values of the area $A$ and circumference
$C$ of the texture mismatch. For both experiment and simulations, we estimated the stiffness $k$ and observe that it increases
with the mismatch between lock and key. The variation of $k$ in experiments and simulations closely match (Fig.~\mvh{\ref{fig:barcodedetails}}c).
Neither $A$ nor $C$ are good predictors for the level of frustration.
To interpret the outcome of the experiments with the $5\times5\times5$ cube, we posit
that for incompatible key textures,
different parts of the cube approach compatible configurations with opposite parity, thus localizing the frustration along internal domain walls. Hence, a single
misplaced pixel carries an energy penalty due to the four frustrated $x$ and $y$-sides of the brick in front of the defect, and one $z$ side
opposing the defect - when defects touch, their interface is not frustrated. Therefore, the size of the domain walls equals the number of frustrated sides, which equals $A+C$.

\clearpage
\section{Extended Data}
\setcounter{figure}{0}
\renewcommand{\figurename}{Extended Data Figure}
\renewcommand\thefigure{ED\arabic{figure}} 

\setcounter{table}{0}
\renewcommand{\tablename}{Extended Data Table}
\renewcommand\thetable{EDT\arabic{table}}

\begin{figure}[h!]
\centerline{\includegraphics[width=1.\columnwidth,clip,trim=2.65cm 6.85cm 6.5cm 7.6cm]{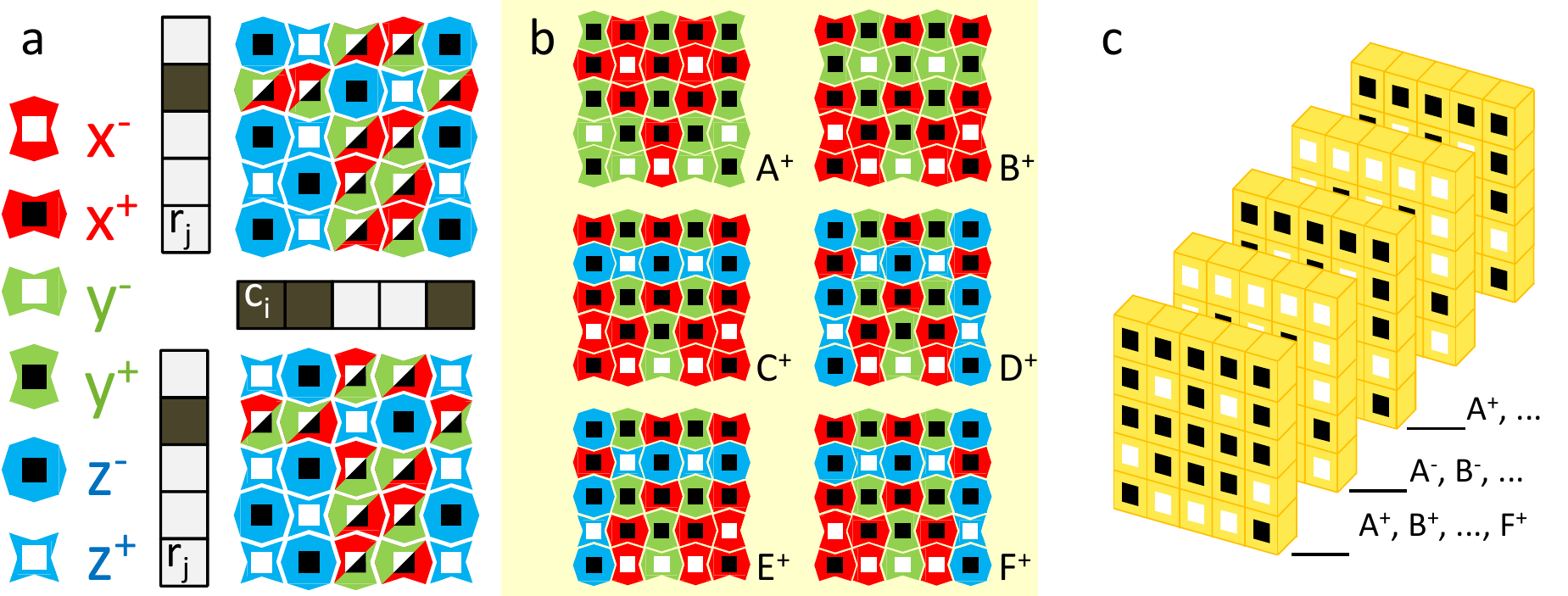}}
\caption{\mvh{Motif Based Design} (a) 2D representation of the six bricks, and illustration
of complex motifs. All complex motifs can be generated by defining two binary vectors $\{c_i\}$ (column) and $\{r_j\}$ (row) that first govern the
placement of $z$-bricks at location $(i,j)$. The remaining sites are filled with
$x$ and $y$-bricks. Respecting parity, this generates all motifs for given $c$ and $r$. (b) The six motifs that are compatible with a $5\times5$
smiley texture (c) A total of $6^5$ smiley metacubes can be designed by varying the stacking order - here $A^-$ denotes the same motif as $A^+$ but with inversed spins. The $x$ and $y$-spins follow from the choice of motifs.
 }\label{fig:combi}
\end{figure}

\begin{figure}[h!]
\centerline{\includegraphics[width=.99\columnwidth]{ExtendedDataFigure2}}
\caption{ \mvh{ Implementation (a)} Computer assisted design of the geometry of the unit cell and (b) a $5\times 5\times 5$ cube.
All our samples were 3D printed with the dimensions $a=11.46$~mm, $D=10.92$~mm, $w=3.6$~mm. To make the wall thickness outside the cube equal
to the internal wall thickness, the outer walls are thickened by $0.27$~mm.}\label{fig:geometry}
\end{figure}

\begin{figure}[h!]
\centerline{\includegraphics[width=1.\columnwidth]{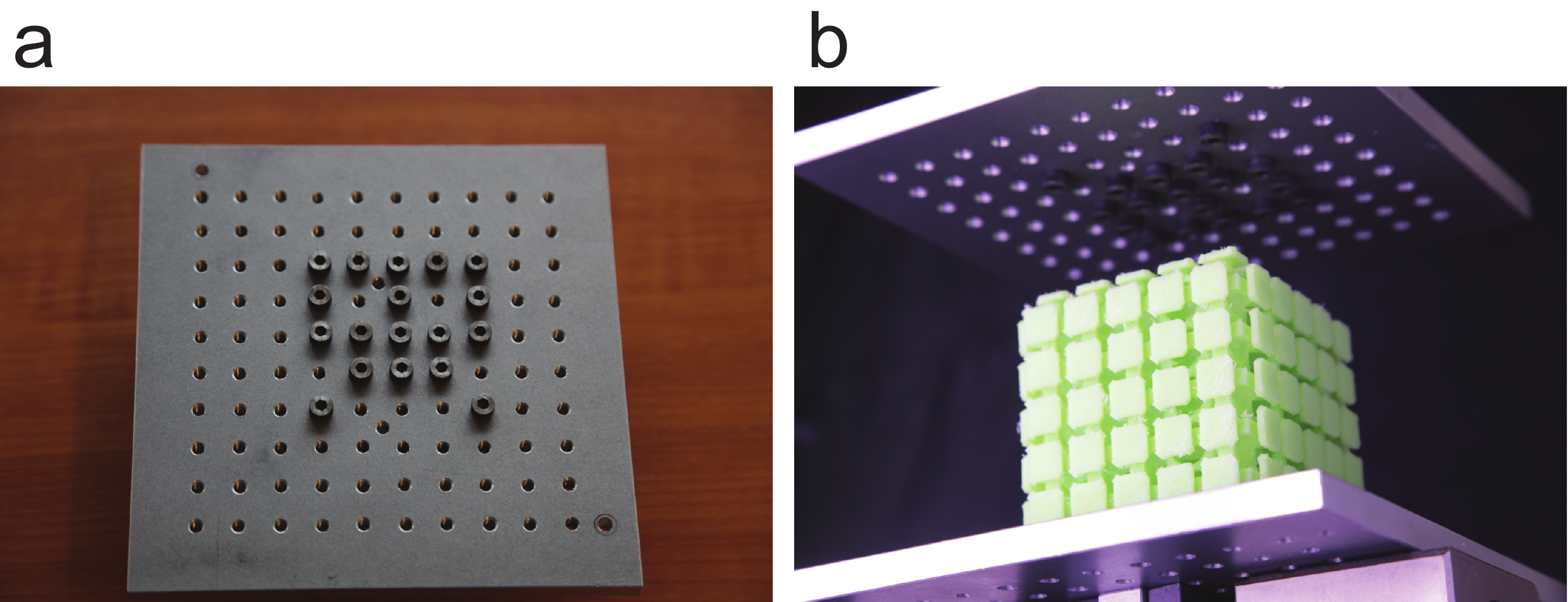}}
\caption{\mvh{ Lock and Key Experiment. (a) Picture of the textured clamp. (b) Side view of the experiment.}}
\label{fig:clamp}
\end{figure}

\begin{figure}[h!]
\centerline{\includegraphics[width=.99\columnwidth,clip,trim=0cm 2.cm 0cm 0cm]{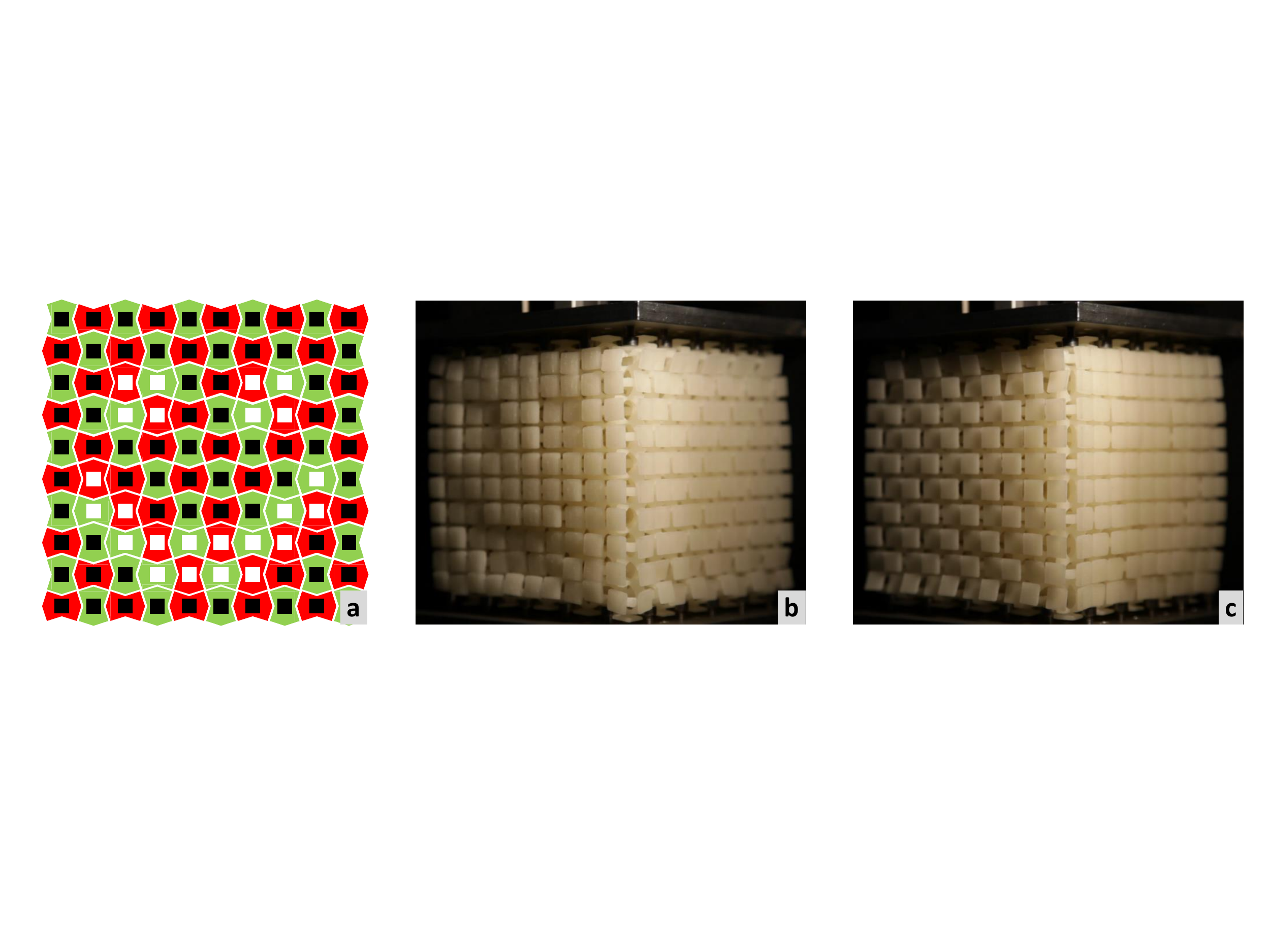}}
\caption{ $10\times10\times10$ metacube under uniaxial compression. (a) Motif $A^+$ - the cube is designed by stacking motifs $A^+$ and $A^-$.
(b) Opposite face of the one shown in Fig. 1e showing the inverted pattern.
(c) One of the transverse faces showing a checkerboard pattern. }\label{fig:antismiley}
\end{figure}

\begin{figure}[h!]
\centerline{\includegraphics[width=1.\columnwidth]{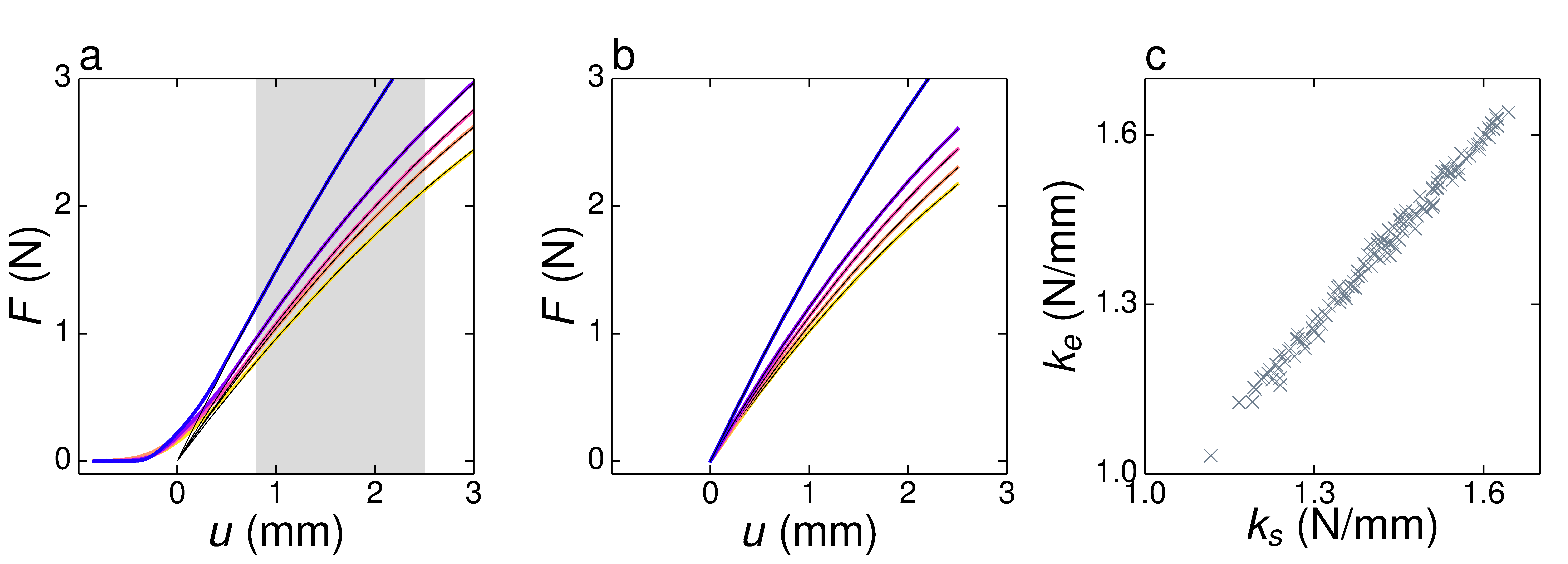}}
\caption{ Complex sensory properties of a complex $5\times 5 \times 5$ metacube with internal smiley texture.
(a) Force-compression curve for five experiments (thick solid lines), where the colour indicates the external
texture shown in  Fig. 3.
The black thin lines show fits to a quadratic function $F_{fit}(u)=k u +\alpha u^2$ performed in the shaded region
$0.8~ \textrm{mm}\le u\le 2.5 ~\textrm{mm}$.
(b) Corresponding numerical results.
(c) Scatter plot showing very good correspondence between the stiffness obtained by simulations ($k_s$) and
experiments ($k_e$).}
\label{fig:barcodedetails}
\end{figure}

\begin{table}[h!]
\centerline{\includegraphics[width=.99\columnwidth]{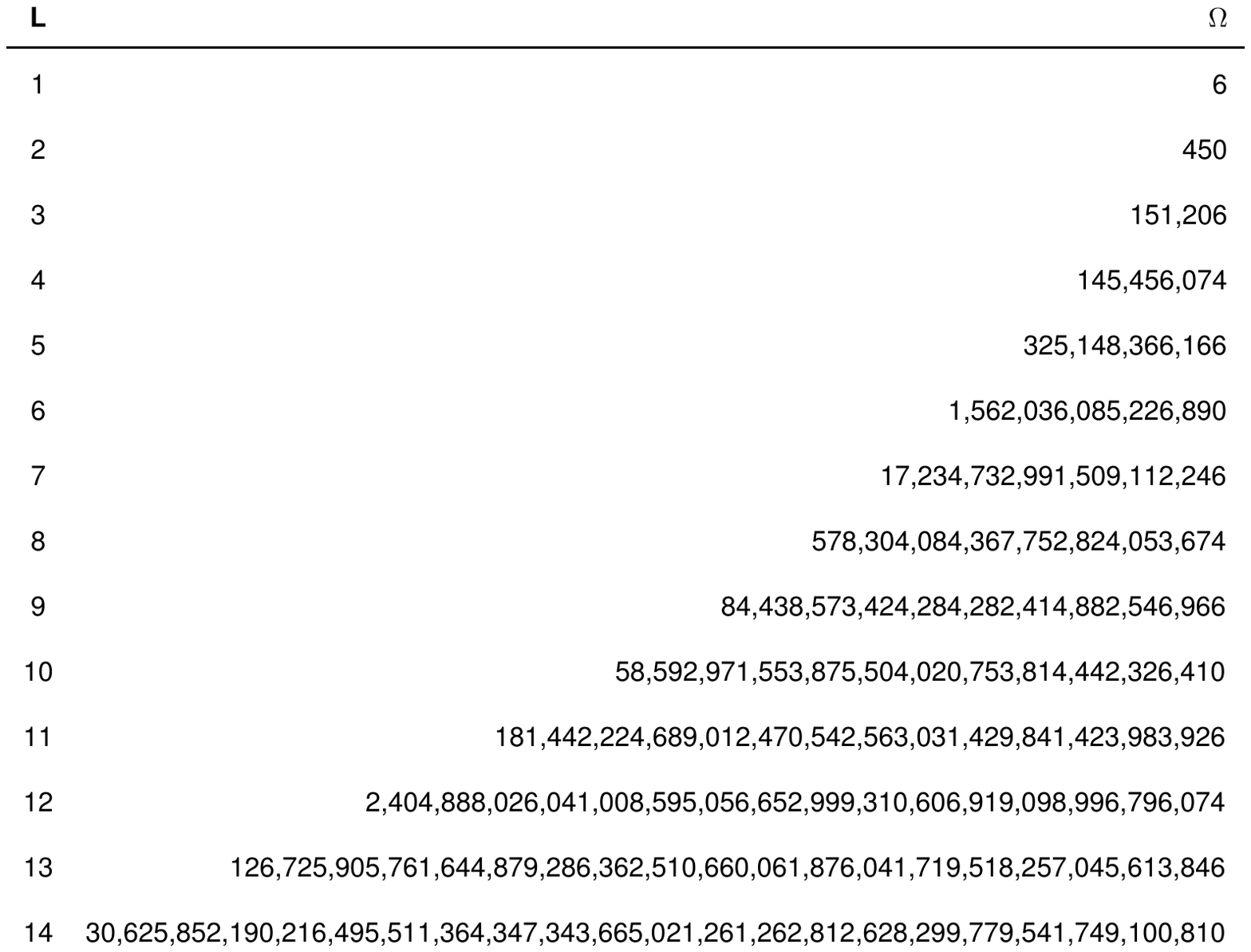}}
\label{tab:omtable}
\caption{The exact value of $\Omega$ for $L\times L \times L$ metacubes up to $L=14$.}
\end{table}

\cleardoublepage
\section*{Supplementary Information}
\setcounter{section}{0}

\section{Movies}
We provide details for the 4 Movies accompanying the main text.
\begin{itemize}
\item The movie \emph{5x5x5\_Periodic.mp4} shows a $5\times5\times 5$ metacube, which is uniaxially
compressed along its minor axis by flat clamps. As discussed in the main text, it exhibits a pattern transformation, where
its building blocks suddenly morph into alternated bricks of elongated and flattened shape.
\item The movie \emph{smiley.MOV} shows a $10\times 10\times 10$ metacube decorated with square pedestals, which is uniaxially
compressed along its minor axis by clamps textured in a checkerboard pattern (see methods). As discussed in the main text,
its surface texture morphs into an exactingly designed "smiley" pattern.
\item The movie \emph{antismiley.MOV} shows the opposite face of the same $10\times 10\times 10$ metacube during a similar
experiment. As discussed in the methods, its surface texture morphs into the inverted "smiley" pattern.
\item The movie \emph{checkerboard.MOV} shows a side face of the same $10\times 10\times 10$ metacube during a similar
experiment. As discussed in the methods, its surface texture morphs into a checkerboard pattern.
\end{itemize}

\section{Combinatorics}
\setcounter{figure}{0}
\renewcommand{\figurename}{SI Figure}
\renewcommand\thefigure{SI\arabic{figure}} 

\setcounter{table}{0}
\renewcommand{\tablename}{SI Table}
\renewcommand\thetable{SIT\arabic{table}} 

\setcounter{equation}{0}
\renewcommand\theequation{SI\arabic{equation}} 

Here we derive a formula for calculating the number of compatible $L \times L \times L$ spin configurations, $\Omega(L)$. In section \ref{bounds} we find lower and upper bounds on $\Omega(L)$ for any $L$. Section \ref{proofs} contains several proofs needed for the derivation of these bounds. Section \ref{exact} contains a detailed derivation of a recurrence equation, which enables us to numerically evaluate $\Omega(L)$ exactly up to $L=14$. In section \ref{estimates} we use the exact results of section \ref{exact} to find an approximate upper bound on $\Omega(L)$.

\begin{figure}
\includegraphics[width=\columnwidth]{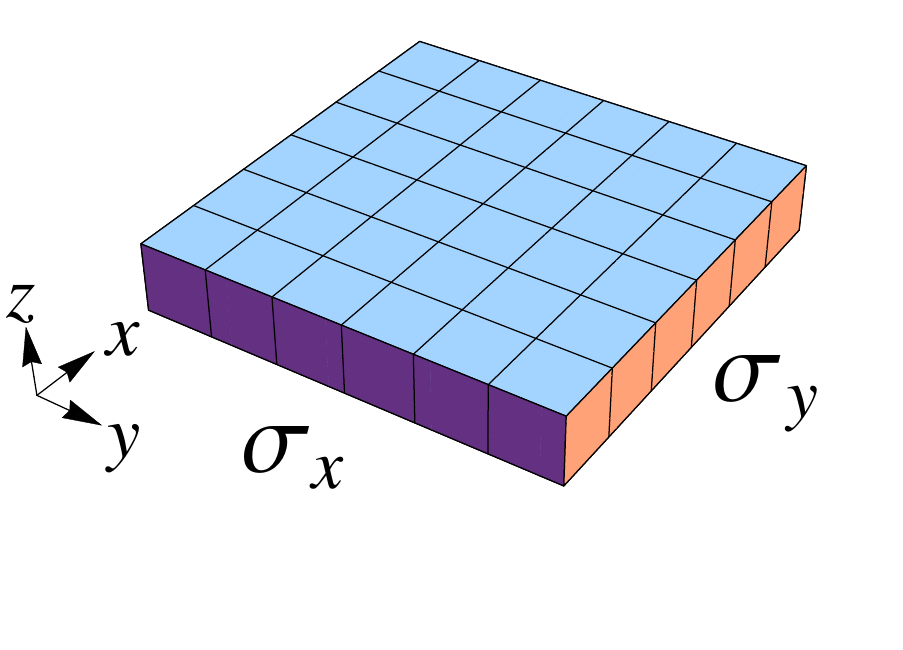}
\caption{An illustration of a $6\times 6\times 1$ configuration. The $\sigma_{z}$ spins are specified on the top $6\times 6$ squares, the $\sigma_{x}$ spins are specified on the row at the front, and the $\sigma_{y}$ spins are specified on the column on the right.}
\label{sigxyz}
\end{figure}

We first consider $L\times L\times 1$ configurations, and say that a row of $L$ spins $\left\{\sigma_{x}\right\}$ and a column of $L$ spins $\left\{\sigma_{y}\right\}$ is a \emph{solution} of a $\left\{\sigma_{z}\right\}$ $L\times L$ texture if the combination of $\left\{\sigma_{x},\sigma_{y},\sigma_{z}\right\}$ yields a compatible $L\times L\times 1$ configuration, see Fig. \ref{sigxyz}. We denote by $Q$ the number of solutions of a given texture $\left\{\sigma_{z}\right\}$, which is equal to the number of corresponding compatible $L\times L\times 1$ configurations. These $L\times L\times 1$ configurations can be stacked in any order, yielding $Q^{L}$ distinct $L\times L\times L$ metacubes for this particular texture, and thus
\begin{align}
\Omega(L)=\sum_{Q}Z_{Q}(L)Q^{L} .\label{omdef}
\end{align}
where $Z_{Q}(L)$ is the number of $\left\{\sigma_{z}\right\}$ $L\times L$ textures that have exactly $Q$ $\left\{\sigma_{x},\sigma_{y}\right\}$ solutions. The number of compatible structures, or distinct metacubes is $\Omega(L)/2$ since each compatible structure has two compatible deformations that differ by a global spin flip.

For simplicity, we define the staggered spins $\tilde{\sigma}$, such that for site $(i,j,k)$ in the metacube
\begin{align}
\tilde{\sigma}^{(i,j,k)}_{\alpha}=\left(-1\right)^{i+j+k}\sigma^{(i,j,k)}_{\alpha} ,
\end{align}
for $\alpha=x,y,z$. One symmetry property that will be used repeatedly is the fact that any two $\tilde{\sigma_{z}}$ textures which differ only by permutations of rows or columns have exactly the same number of $\left\{\tilde{\sigma}_{x},\tilde{\sigma_{y}}\right\}$ solutions, namely they have the same value of $Q$. Therefore, if a texture contains $p$ columns which are all $+1$, we may assume without loss of generality that these columns are the leftmost columns, and this particular choice of placing these columns represents $\left(\begin{array}{c}L\\p\end{array}\right)$ textures with the same value of $Q$.

\subsection{Bounds on $\Omega$} \label{bounds}

We now find lower and upper bounds on $\Omega(L)$ by bounding $Z_{Q}$. First note that each $\tilde{\sigma}_{z}$ texture has at least two solutions, i.e. $Q\geq2$ for all textures; one in which all the $\tilde{\sigma}_{x}$ spins are equal to $+1$ and the $\tilde{\sigma}_{y}$ spins are equal to $-1$, and one in which all the $\tilde{\sigma}_{x}$ spins are equal to $-1$ and the $\tilde{\sigma}_{y}$ spins are equal to $+1$. A simple lower bound on $\Omega$ is found by saying that all textures have at least $Q=2$. Since there are in total $2^{L^{2}}$ $\tilde{\sigma}_{z}$ textures, we find that
\begin{align}
\Omega(L)\geq 2^{L^{2}+L} .\label{simplelower}
\end{align}

\subsubsection{Simple bounds}

For a simple upper bound we note that there are exactly two textures that have the maximal number of solutions: $\tilde{\sigma}_{z}\equiv+1$ and $\tilde{\sigma}_{z}\equiv-1$. Each of these textures has $Q=2^{L+1}-1$ solutions; Consider for example $\tilde{\sigma}_{z}\equiv+1$. If $\tilde{\sigma}_{x}\equiv-1$ then all the $\tilde{\sigma}_{y}$ are free, leading to $2^{L}$ solutions, if $\tilde{\sigma}_{y}\equiv-1$ then all the $\tilde{\sigma}_{x}$ are free, leading to an additional $2^{L}$ solutions, however $\tilde{\sigma}_{x}\equiv\tilde{\sigma}_{y}\equiv-1$ was counted twice, and thus the total number of solutions is $2^{L}+2^{L}-1=2^{L+1}-1$. Below we will show that all other $\tilde{\sigma}_{z}$ textures have less solutions.

The next highest number of solutions for a given $\tilde{\sigma}_{z}$ $Q=2^{L}+2^{L-1}=3\cdot2^{L-1}$ is for $\tilde{\sigma}_{z}$ textures which are all $+1$ (or $-1$) except for one row or column which is all $-1$ (or $+1$, respectively). See proof in section \ref{proof2} below.

A lower bound on the number of compatible configurations for a $L \times L \times L$ cube is obtained by saying that except for the two $\tilde{\sigma}_z$ textures with the maximal number $2^{L+1}-1$ of solutions, all the other $2^{L^2}-2$ textures have \emph{at least} two solutions, and thus
\bea
\Omega(L) \ge \left[2\cdot \left( 2^{L+1} - 1 \right)^L + \left( 2^{L^2} - 2 \right) \cdot 2^L\right] . \label{eq:lower_exact}
\eea
For $L \gg 1$, this may be approximated by
\bea
\Omega(L) \gtrsim 3 \cdot 2^{L^2+L} . \label{eq:lower_asymp}
\eea
Note that this result may also be obtained by considering the lower bound given in Eq. (\ref{simplelower}) above and applying the arguments leading to it on all three directions. For $L \gg 1$ the multiple counting of the same configuration from different directions is expected to be negligible.

For the upper bound, we say that except for the two $\tilde{\sigma}_z$ textures that have the maximal number $2^{L+1}-1$ of solutions, all the other $2^{L^2}-2$ textures have \emph{at most} $2^L + 2^{L-1} = 3 \cdot 2^{L-1}$ solutions, and thus 
\bea
\Omega(L) &\le& 2 \cdot \left( 2^{L+1} - 1 \right)^L + \left( 2^{L^2} - 2 \right) \cdot \left(3\cdot2^{L-1} \right)^L . \label{eq:upper_exact}
\eea

For $L \gg 1$, this may be approximated by
\bea
\Omega(L) \lesssim 2^{2 L^2} \left( \frac{3}{2} \right)^L .\label{eq:upper_asymp}
\eea

\subsubsection{Better upper bound}

\begin{figure}
\includegraphics[width=\columnwidth]{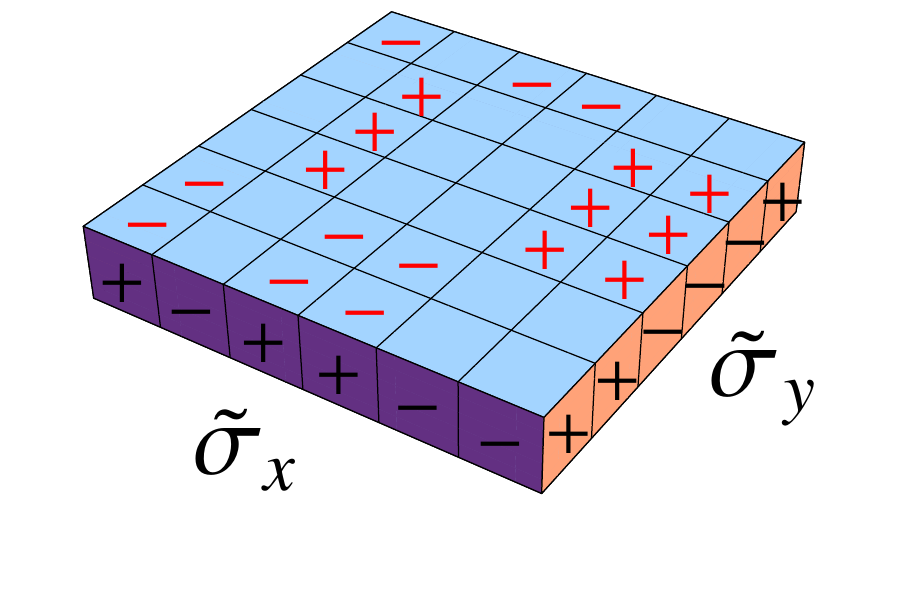}
\caption{An example of a $6\times6\times1$ configuration. $p_{x}=3$ of the $\tilde{\sigma}_{x}$ spins are $+1$, and $p_{y}=3$ of the $\tilde{\sigma}_{y}$ spins are $+1$. The $\tilde{\sigma}_{z}$ spins in the intersections between $+1$ ($-1$) $\tilde{\sigma}_{x}$ and $\tilde{\sigma}_{y}$ spins must be $-1$ ($+1$).}
\label{ex1}
\end{figure}

An even better upper bound can be found by finding a lower bound on $Z_{2}(L)$, the number of $\tilde{\sigma}_{z}$ textures that have only the two trivial $\tilde{\sigma}_{x}-\tilde{\sigma}_{y}$ solutions. Consider a $\tilde{\sigma}_{z}$ that has a solution in which $p_{x}$ of its $\tilde{\sigma}_{x}$ spins are in the $+1$ state and $p_{y}$ of its $\tilde{\sigma}_{y}$ spins are in the $+1$ state. In that case, the $\tilde{\sigma}_{z}$ spins in the intersection between the $p_{x}$ $\tilde{\sigma}_{x}=+1$ and $p_{y}$ $\tilde{\sigma}_{y}=+1$ spins must be $\tilde{\sigma}_{z}=-1$, and in the intersection between the $L-p_{x}$ $\tilde{\sigma}_{x}=-1$ and $L-p_{y}$ $\tilde{\sigma}_{y}-1$ spins must be $\tilde{\sigma}_{z}=+1$, see Fig. \ref{ex1}. The other $\left(L-p_{x}\right)p_{y}+\left(L-p_{y}\right)p_{x}$ spins are in the intersection of opposite values of $\tilde{\sigma}_x$ and $\tilde{\sigma}_y$, and thus are free to be $\tilde{\sigma}_{z}=\pm1$. Therefore, the number of $\tilde{\sigma}_{z}$ textures that have a solution with these values of $p_{x}$ and $p_{y}$ $+1$ spins in $\tilde{\sigma}_{x}$ and $\tilde{\sigma}_{y}$ respectively is
\begin{align}
M\left(p_{x},p_{y}\right)=\left(\begin{array}{c}L\\p_{x}\end{array}\right)\left(\begin{array}{c}L\\p_{y}\end{array}\right)2^{\left(L-p_{x}\right)p_{y}+\left(L-p_{y}\right)p_{x}} .
\end{align}

A lower bound on $Z_{2}(L)$ can be found by excluding all the $\tilde{\sigma}_z$ textures that have more than two solutions. A bound on that may be obtained by summing over $M\left(p_{x},p_{y}\right)$ and noting that in this way each $\tilde{\sigma}_z$ texture is counted at least once,
\begin{align}
&Z_{2}(L)\geq\nonumber\\
&\geq2^{L^{2}}-\left[\sum^{L}_{p_{x}}\sum^{L}_{p_{y}}M\left(p_{x},p_{y}\right)-M\left(0,L\right)-M\left(L,0\right)\right]=\nonumber\\
&=3\cdot2^{L^{2}}-\sum^{L}_{p=0}\left(\begin{array}{c}L\\p\end{array}\right)\left(2^{p}+2^{L-p}\right)^{L} .
\label{minz2}
\end{align}
At large $L$, the main contribution to the sum comes from the extreme values of $p$, $p\approx0$ and $p\approx L$. Therefore, as an approximation we include in the sum only the terms $p=0,1,L-1,L$ and find that
\begin{align}
&Z_{2}(L)\geq2^{L^{2}}\left(1-\frac{4L}{2^{L}}\right) .
\end{align}
To obtain an upper bound on $\Omega(L)$, we say that the number of $\tilde{\sigma}_{z}$ textures that have two solutions is at least that given by Eq. (\ref{minz2}), two of the textures have $2^{L+1}-1$ solutions, and the rest have $3\cdot2^{L-1}$ solutions, such that
\begin{align}
&\Omega\left( L \right) \leq \left[3 \cdot 2^{L^{2}} - \sum^{L}_{p=0} \left(\begin{array}{c}L\\p\end{array}\right) \left(2^{p}+2^{L-p}\right)^{L} \right] 2^{L}+ \nonumber\\
&+2\cdot\left(2^{L+1}-1\right)^{L}+\nonumber\\
&+\left[\sum^{L}_{p=0}\left(\begin{array}{c}L\\p\end{array}\right)\left(2^{p}+2^{L-p}\right)^{L}-2^{L^{2}+1}-2\right]\left(3\cdot2^{L-1}\right)^{L} \nonumber\\
&=3\cdot 2^{L^{2}+L}-2^{L^{2}-L+1} \cdot 3^{L} \cdot \left(2^{L^{2}}+1 \right) + 2 \cdot \left(2^{L+1}-1\right)^{L}+\nonumber\\
&+\left(2^{L^{2}-L} \cdot 3^{L} - 2^{L} \right) \sum^{L}_{p=0}\left(\begin{array}{c}L\\p\end{array}\right) \left(2^{p}+2^{L-p}\right)^{L} . \label{eq:upper_exact2}
\end{align}
The first line in the first expression in Eq. (\ref{eq:upper_exact2}) corresponds to the $\tilde{\sigma}_{z}$ textures that have two solutions, the second line corresponds to the two textures that have $2^{L+1}-1$ solutions, and the third line corresponds to the other textures. In the limit $L\gg1$, we consider in the sum over $p$ only the terms $p=0,1,L-1,L$, and thus Eq. (\ref{eq:upper_exact2}) may be approximated by
\begin{align}
&\Omega\left(L\right)\leq 4L\cdot2^{2L^{2}}\left(\frac{3}{4}\right)^{L} .\label{eq:upper_asymp2}
\end{align}

In summary, we find that for large $L$, $\Omega(L)$ is bounded by
\begin{align}
&3\cdot2^{L^{2}+L}\leq\Omega\left(L\right)\leq4L\cdot2^{2L^{2}}\left(\frac{3}{4}\right)^{L} .
\end{align}

\subsection{Proofs} \label{proofs}

\subsubsection{Maximal number of compatible configurations} \label{proof1}

Here we prove that for a layer of size $k\times L$, the $\tilde{\sigma}_{z}$ textures that have the maximal number of solutions are $\tilde{\sigma}_z \equiv +1$ and $\tilde{\sigma}_z\equiv -1$, and each of them has $Q=2^{k}+2^{L}-1$. We do this by induction on $L$. 

For $L=1$, assume that the $\tilde{\sigma}_{z}$ texture has $p$ $+1$ spins and $k-p$ $-1$ spins. If the only $\tilde{\sigma}_{y}$ spin is $+1$, the $p$ spins in $\tilde{\sigma}_{x}$ corresponding to the $p$ $\tilde{\sigma}_{z}$ spins which are equal to $+1$ must be $-1$, and the remaining $k-p$ $\tilde{\sigma}_{x}$ spins are free. Similarly, if the $\tilde{\sigma}_{y}$ spin is $-1$, then $p$ of the $\tilde{\sigma}_{x}$ spins are free and the rest must be $+1$. Therefore, a texture of size $k\times 1$ with $p$ $+1$ spins has $Q=2^{p}+2^{k-p}$ solutions. This is maximal when either $p=0$ or $p=k$, and then $Q=2^{k}+1=2^{k}+2^{L}-1$, as required.

\begin{figure}
\subfigure[]{\includegraphics[width=0.3\columnwidth]{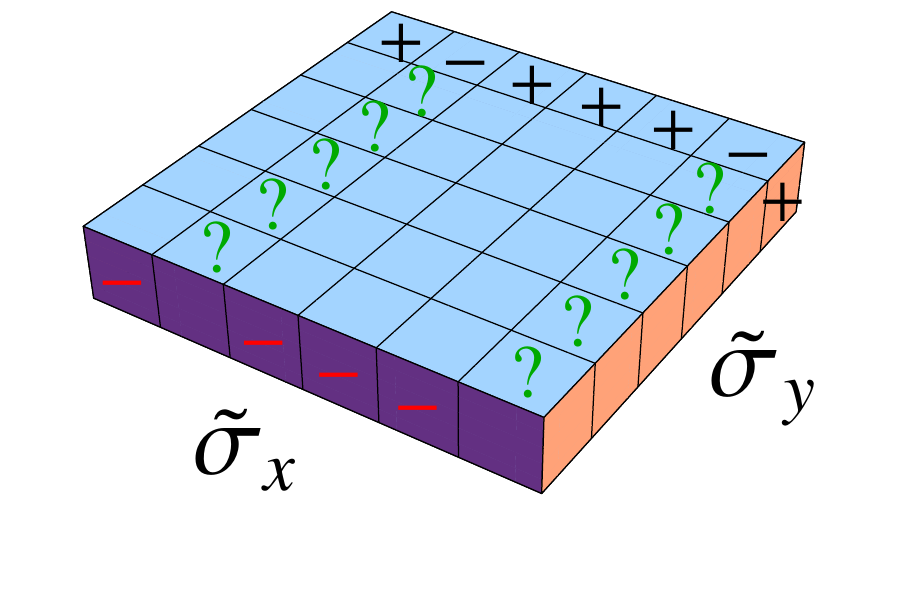}}
\subfigure[]{\includegraphics[width=0.3\columnwidth]{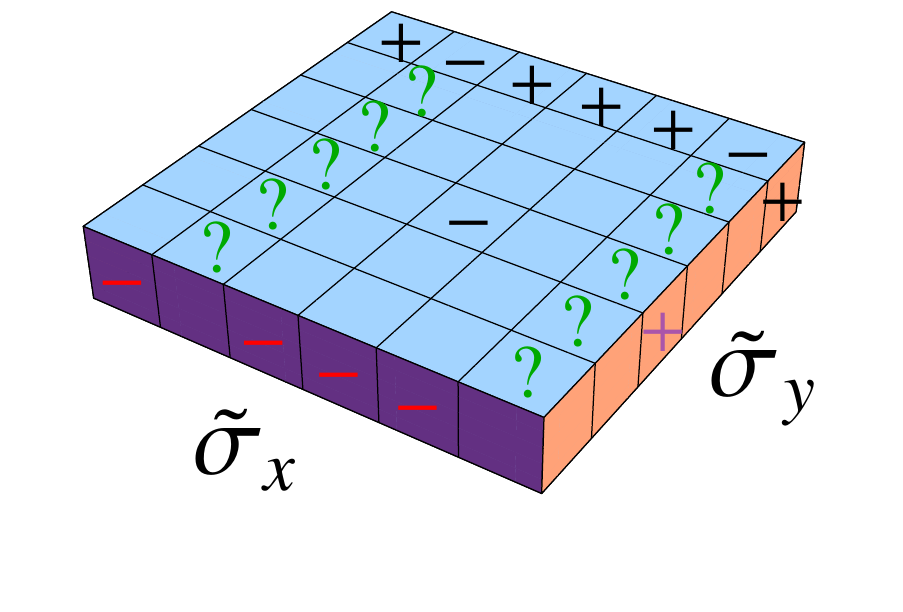}}
\subfigure[]{\includegraphics[width=0.3\columnwidth]{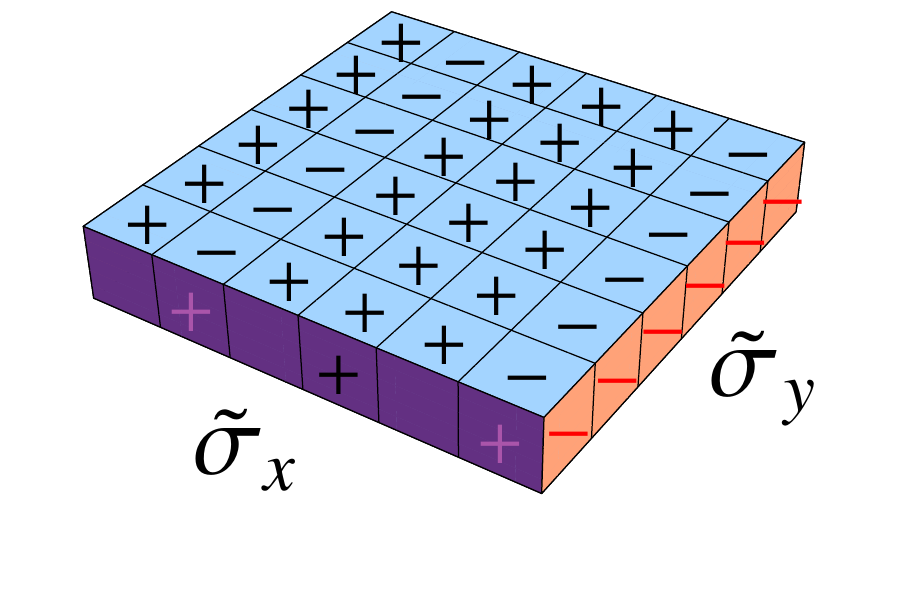}}
\caption{(a) If the top row has $p=4$ $+1$ $\tilde{\sigma}_{z}$ spins and the top $\tilde{\sigma}_{y}$ spin is $+1$, then the $p=4$ corresponding $\tilde{\sigma}_{x}$ spins are $-1$. The other $\tilde{\sigma}_{x}$ and $\tilde{\sigma}_{y}$ spins are determined by at least the $\tilde{\sigma}_{z}$ spins marked with a green question mark. (b) If not all of the columns are homogeneous, then the remaining $\tilde{\sigma}_{x}$ and $\tilde{\sigma}_{y}$ spins are not determined solely by the $\tilde{\sigma}_{z}$ spins marked with a green question mark. (c) If at least one of the $p=4$ $\tilde{\sigma}_{x}$ spins corresponding to the all $+1$ columns is $+1$ (in black), then all the (red) $\tilde{\sigma}_{y}$ spins are $-1$, and subsequently the $L_{x}-p_{x}=2$ $\tilde{\sigma}_{x}$ (purple) spins are $+1$.}
\label{proof1fig}
\end{figure}

Now we will assume this is true for all sizes until $L$ and will check for $L+1$. Assume that the top row in the $\tilde{\sigma}_{z}$ texture has $p$ $+1$ spins and $k-p$ $-1$ spins, see Fig. \ref{proof1fig}a. If the top $\tilde{\sigma}_{y}$ spin is $+1$, then $p$ of the $\tilde{\sigma}_{x}$ spins must be $-1$, and from the other restrictions there are at most $2^{L}+2^{k-p}-1$ solutions (according to the induction assumption), and similarly if the top $\tilde{\sigma}_{y}$ spin is $-1$. In total, there are at most $2^{L}+2^{k-p}-1+2^{L}+2^{p}-1$ solutions, which is maximal for either $p=0$ or $p=k$, and in those cases we get the required result.

\subsubsection{Second maximal number of compatible configurations} \label{proof2}

Here we prove that for a layer of size $L\times L$, the $\tilde{\sigma}_{z}$ textures with the second maximal number of solutions are those that are all $+1$ ($-1$), except for a single row or column which is all $-1$ ($+1$). The number of solutions for this type of texture is $Q=2^{L-1}+2^{L}$.

Let us consider a $\tilde{\sigma}_{z}$ which is not all $+1$ or all $-1$. Without loss of generality, we may assume that at least one of the rows is not all $+1$ or all $-1$ (if all the rows are such, we rotate the texture by $90$ degrees). Let's assume that this row has $0<p<L$ $+1$ spins and $0<L-p<L$ $-1$ spins, and without loss of generality we may assume that this is the top row, see Fig. \ref{proof1fig}a. If the top $\tilde{\sigma}_{y}$ spin is $+1$, then the $p$ $\tilde{\sigma}_{x}$ spins corresponding to the $p$ $+1$ spins in the top row must be $-1$. If all the $p$ columns are all $+1$ there are no further constraints on the remaining $L-p$ $\tilde{\sigma}_{x}$ spins and on the remaining $L-1$ $\tilde{\sigma}_{y}$ spins, see Fig. \ref{proof1fig}b. A similar argument can be made if the top $\tilde{\sigma}_{y}$ spin is $-1$. Hence, if the top $\tilde{\sigma}_{z}$ row has $p$ $+1$ spin and $L-p$ $-1$ spin, the maximal number of solutions is obtained when each column is either all $+1$ or all $-1$.

If at least one of the $\tilde{\sigma}_{x}$ spins corresponding to the $p$ $+1$ columns is $+1$, then all the $\tilde{\sigma}_{y}$ spins must be equal to $-1$, and thus the remaining $L-p$ $\tilde{\sigma}_{x}$ spins must be equal to $+1$, see Fig. \ref{proof1fig}c. Similarly, if at least one of the $\tilde{\sigma}_{z}$ spins corresponding to the $L-p$ $-1$ columns is $-1$, then all the $\tilde{\sigma}_{y}$ spins must be equal to $+1$, and thus the remaining $p$ $\tilde{\sigma}_{x}$ spins must be equal to $-1$. Otherwise, the $\tilde{\sigma}_{y}$ spins are free. Hence, in total there are $Q=2^{p}-1+2^{L-p}-1+2^{L}$ solutions. Since $p\neq0,L$, the maximum is obtained when $p=1$ or $p=L-1$, in which case $Q=2^{L-1}+2^{L}$.

\subsection{Exact derivation of $Z_{Q}$} \label{exact}

Here we derive an exact recursion relation on $Z_{Q}(L)$ which in principle may be solved numerically for any finite $L$. Combined with Eq. (\ref{omdef}), this yields an exact result for the number of compatible spin configurations $\Omega(L)$.

We divide all $\tilde{\sigma}_z$ textures to the following types::\\
Type $0$ textures do not have any full rows or columns (full means that all the spins in it are the same).\\
Type $C_{\pm}(p)$ textures have $p\geq1$ columns of all $\pm1$ spins, no columns of all $\mp1$ spins, and no rows of all $\pm1$ spins.\\
Type $R_{\pm}(p)$ textures have $p\geq1$ rows of all $\pm1$ spins, no rows of all $\mp1$ spins, and no columns of all $\pm1$ spins.\\
Type $C(p_{+},p_{-})$ textures have $p_{+}\geq1$ columns of all $+1$ spins, and $p_{-}\geq1$ columns of all $-1$ spins.\\
Type $R(p_{+},p_{-})$ textures have $p_{+}\geq1$ rows of all $+1$ spins, and $p_{-}\geq1$ rows of all $-1$ spins.\\
Type $CR_{\pm}(p_{x},p_{y})$ textures have $p_{x}\geq1$ columns of all $\pm1$ spins and $p_{y}\geq1$ rows of all $\pm1$ spins.

We also define $Z^{\beta}_{Q}\left(L_{x},L_{y}\right)$ as the number of $L_{x}\times L_{y}$ textures of type $\beta$ that have $Q$ solutions. From symmetry we have
\begin{align}
&Z^{0}_{Q}\left(L_{x},L_{y}\right)=Z^{0}_{Q}\left(L_{y},L_{x}\right) ,\nonumber\\
&Z^{C_{+}(p)}_{Q}\left(L_{x},L_{y}\right)=Z^{C_{-}(p)}_{Q}\left(L_{x},L_{y}\right)=\nonumber\\
&=Z^{R_{+}(p)}_{Q}\left(L_{y},L_{x}\right)=Z^{R_{-}(p)}_{Q}\left(L_{y},L_{x}\right) ,\nonumber\\
&Z^{C(p_{+},p_{-})}_{Q}\left(L_{x},L_{y}\right)=Z^{C(p_{-},p_{+})}_{Q}\left(L_{x},L_{y}\right)=\nonumber\\
&=Z^{R(p_{+},p_{-})}_{Q}\left(L_{y},L_{x}\right)=Z^{R(p_{-},p_{+})}_{Q}\left(L_{y},L_{x}\right) ,\nonumber\\
&Z^{CR_{+}(p_{x},p_{y})}_{Q}\left(L_{x},L_{y}\right)=Z^{CR_{-}(p_{x},p_{y})}_{Q}\left(L_{x},L_{y}\right)=\nonumber\\
&=Z^{CR_{+}(p_{y},p_{x})}_{Q}\left(L_{y},L_{x}\right)=Z^{CR_{-}(p_{y},p_{x})}_{Q}\left(L_{y},L_{x}\right) .
\end{align}

We now consider a texture of each type which has $Q$ solutions, and see how it can be built from smaller textures which have $q$ solutions. This yields a set of recurrence equations which relate $Z^{\beta}_{Q}\left(L_{x},L_{y}\right)$ to $Z^{\beta'}_{q}\left(\ell_{x},\ell_{y}\right)$ with $q<Q,\ell_{x}<L_{x},\ell_{y}<L_{y}$. The final set of recurrence equations is summarized in Section \ref{finalresult}.

\subsubsection{Types $CR_{\pm}(L_{x},L_{y})$}
In this case, the $\tilde{\sigma}_{z}$ texture is all $+1$ or all $-1$, and $Q=2^{L_{x}}+2^{L_{y}}-1$. Hence,
\begin{align}
Z^{CR_{\pm}(L_{x},L_{y})}_{Q}(L_{x},L_{y})=\delta_{Q,2^{L_{x}}+2^{L_{y}}-1} .
\end{align}

\subsubsection{Types $CR_{\pm}(p_{x},p_{y})$, $1\leq p_{x}\leq L_{x}-1$, $1\leq p_{y}\leq L_{y}-1$}
\begin{figure}
\subfigure[]{\includegraphics[width=0.45\columnwidth]{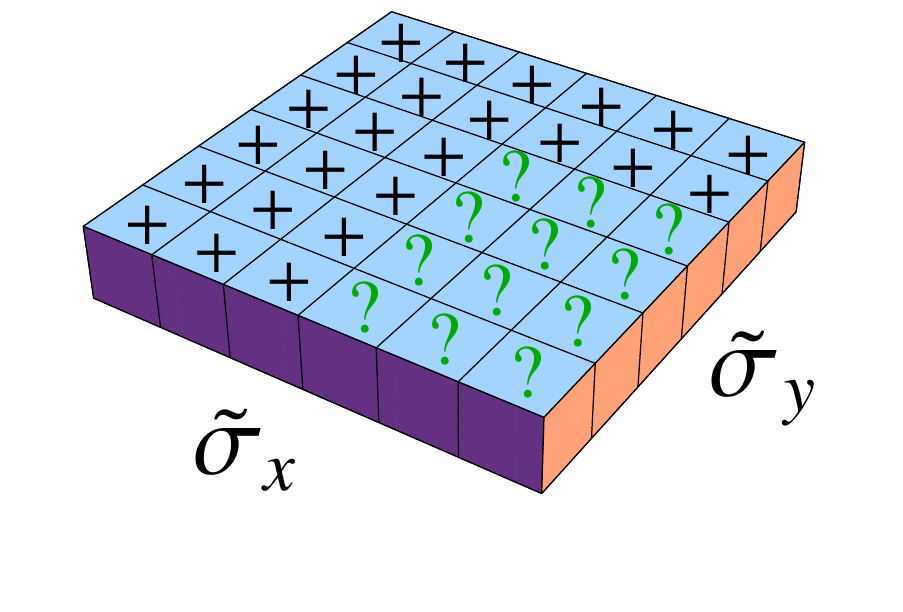}}
\subfigure[]{\includegraphics[width=0.45\columnwidth]{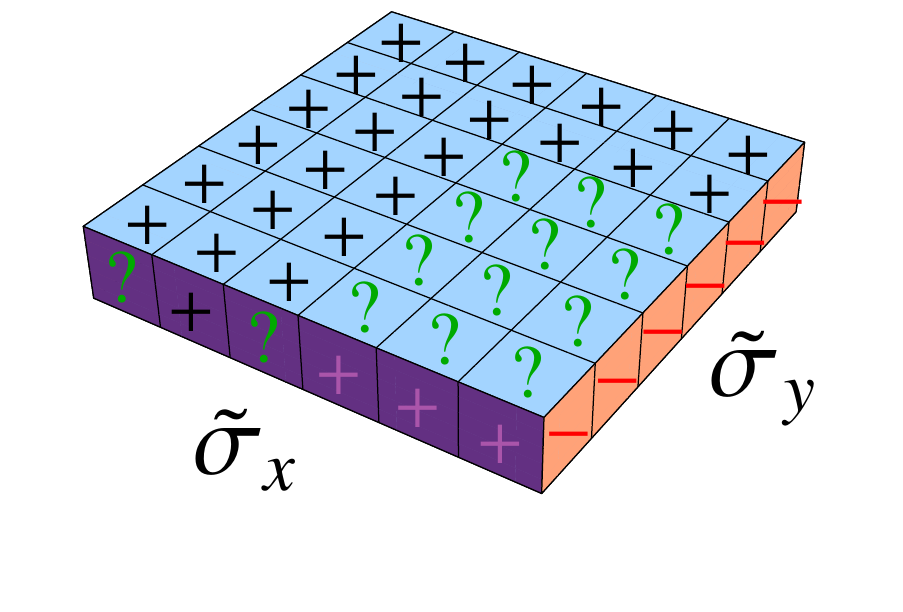}}\\
\subfigure[]{\includegraphics[width=0.45\columnwidth]{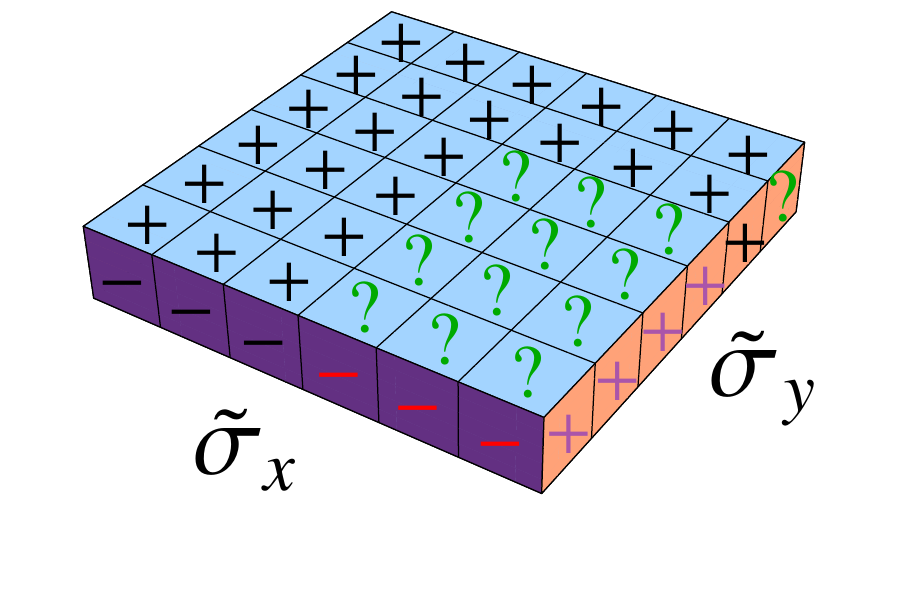}}
\subfigure[]{\includegraphics[width=0.45\columnwidth]{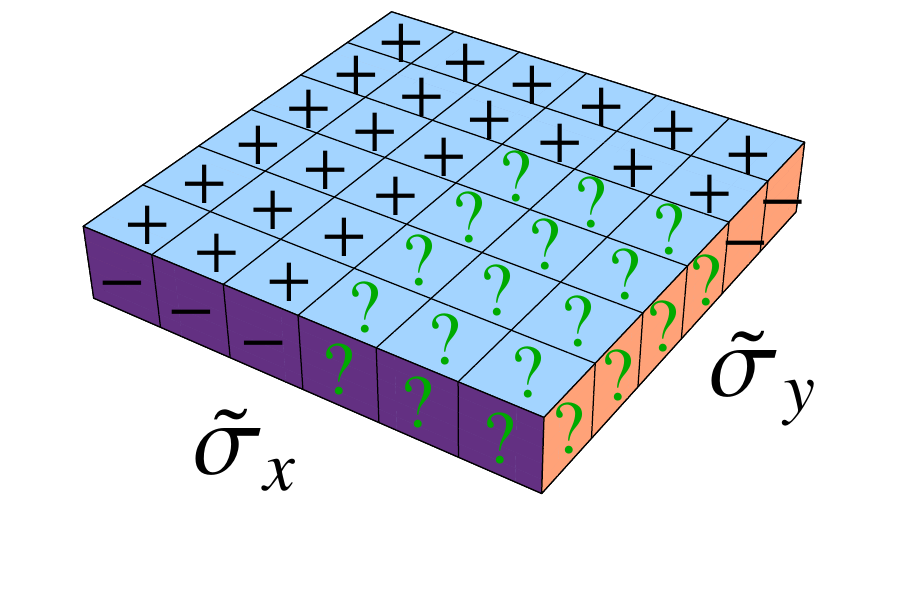}}
\caption{(a) A texture of type $CR_{+}(3,2)$. Each row and column in the $3\times4$ block at the bottom-right corner contains at least one $-1$ spin. (b) If at least one of the leftmost $p_{x}=3$ $\tilde{\sigma}_{x}$ spins is $+1$ (black), the (red) $\tilde{\sigma}_{y}$ spins are $-1$ and subsequently the (purple) rightmost $L_{x}-p_{x}=3$ $\tilde{\sigma}_{x}$ spins are $+1$. (c) If all the leftmost $p_{x}=3$ $\tilde{\sigma}_{x}$ spins are $-1$ and at least one of the $p_{y}=2$ top $\tilde{\sigma}_{y}$ spins (black) is $+1$, the (red) rightmost $L_{x}-p_{x}=3$ $\tilde{\sigma}_{x}$ spins are $-1$. Since each row in the bottom-right block contains at least one $-1$ spin, the bottom $L_{y}-p_{y}$ $\tilde{\sigma}_{y}$ spins (purple) are $+1$. (d) If all the leftmost $p_{x}=3$ $\tilde{\sigma}_{x}$ spins and the top $p_{y}=2$ $\tilde{\sigma}_{y}$ spins are $-1$, the constraints on the remaining $\tilde{\sigma}_{x}$ and $\tilde{\sigma}_{y}$ spins come from the bottom-right block.}
\label{crex}
\end{figure}

We assume without loss of generality that the texture is of type $CR_{+}(p_{x},p_{y})$, that the $p_{x}$ $+1$ columns are the leftmost columns, and that the $p_{y}$ $+1$ rows are the topmost rows. In the bottom right of the texture there is a block of size $(L_{x}-p_{x})\times(L_{y}-p_{y})$, which is a texture of either type $0,C_{-}(k),R_{-}(k)$ or $CR_{-}(k_{x},k_{y})$ with $q$ solutions. In any case, each column and row in the block contains at least one $-1$ spin. See Fig. \ref{crex}a.

If at least one of the leftmost $p_{x}$ $\tilde{\sigma}_{x}$ spins is $+1$, then $\tilde{\sigma}_{y}=-1$. Since each column in the block has at least one $-1$ spin, the $L_{x}-p_{x}$ rightmost $\tilde{\sigma}_{x}$ spins are $+1$. The other leftmost $p_{x}-1$ $\tilde{\sigma}_{x}$ spins are free. This gives a total of $2^{p_{x}}-1$ solutions, see Fig. \ref{crex}b.

If all the leftmost $p_{x}$ $\tilde{\sigma}_{x}$ spins are $-1$, and at least one of the topmost $p_{y}$ $\tilde{\sigma}_{y}$ spins is $+1$, then the rightmost $L_{x}-p_{x}$ $\tilde{\sigma}_{x}$ spins are $-1$. Since each row in the block has at least one $-1$ spin, the bottom $L_{y}-p_{y}$ $\tilde{\sigma}_{y}$ spins are $+1$. The other $p_{y}-1$ topmost $\tilde{\sigma}_{y}$ spins are free. This gives another $2^{p_{y}}-1$ solutions, see Fig. \ref{crex}c.

If the top $p_{y}$ $\tilde{\sigma}_{y}$ spins and the leftmost $p_{x}$ $\tilde{\sigma}_{x}$ spins are $-1$, the only restriction on the remaining $\tilde{\sigma}_{x}$ and $\tilde{\sigma}_{y}$ spins comes from the block, which gives $q$ solutions, see Fig. \ref{crex}d.

Therefore, if the block has $q$ solutions, the $L_{x}\times L_{y}$ texture has $Q=2^{p_{x}}+2^{p_{y}}-2+q$ solutions.

Hence, we find that for $1\leq p_{x}\leq L_{x}-1$ and $1\leq p_{y}\leq L_{y}-1$
\begin{align}
&Z^{CR_{\pm}(p_{x},p_{y})}_{Q}(L_{x},L_{y})=\left(\begin{array}{c}L_{x}\\p_{x}\end{array}\right)\left(\begin{array}{c}L_{y}\\p_{y}\end{array}\right)\delta_{Q,2^{p_{x}}+2^{p_{y}}-2+q}\nonumber\\
&\Biggl[\sum^{L_{x}-p_{x}}_{k_{x}=1}\sum^{L_{y}-p_{y}}_{k_{y}=1}Z^{CR_{\mp}(k_{x},k_{y})}_{q}(L_{x}-p_{x},L_{y}-p_{y})+\nonumber\\
&\left.+\sum^{L_{x}-p_{x}-1}_{k=1}Z^{C_{\mp}(k)}_{q}(L_{x}-p_{x},L_{y}-p_{y})+\right.\nonumber\\
&\left.+\sum^{L_{y}-p_{y}-1}_{k=1}Z^{R_{\mp}(k)}_{q}(L_{x}-p_{x},L_{y}-p_{y})+\right.\nonumber\\
&+Z^{0}_{q}(L_{x}-p_{x},L_{y}-p_{y})\Biggr] .
\end{align}

\subsubsection{Types $C(p,L_{x}-p)$ and $R(p,L_{y}-p)$}
\begin{figure}
\subfigure[]{\includegraphics[width=0.45\columnwidth]{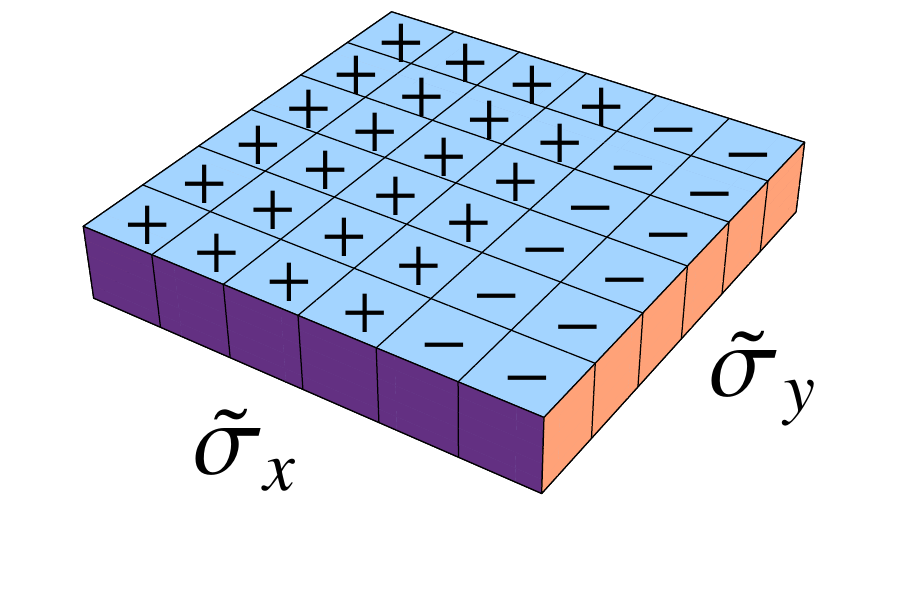}}
\subfigure[]{\includegraphics[width=0.45\columnwidth]{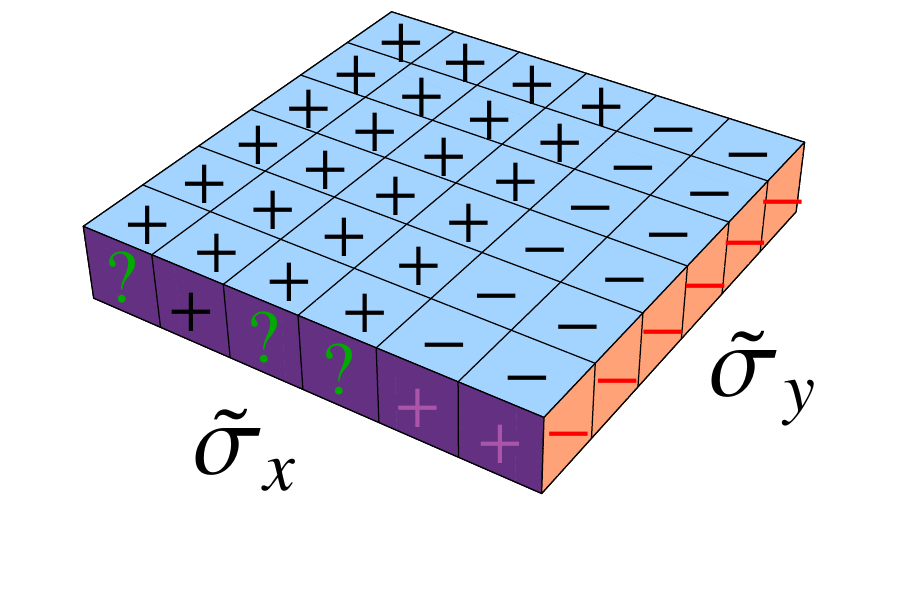}}\\
\subfigure[]{\includegraphics[width=0.45\columnwidth]{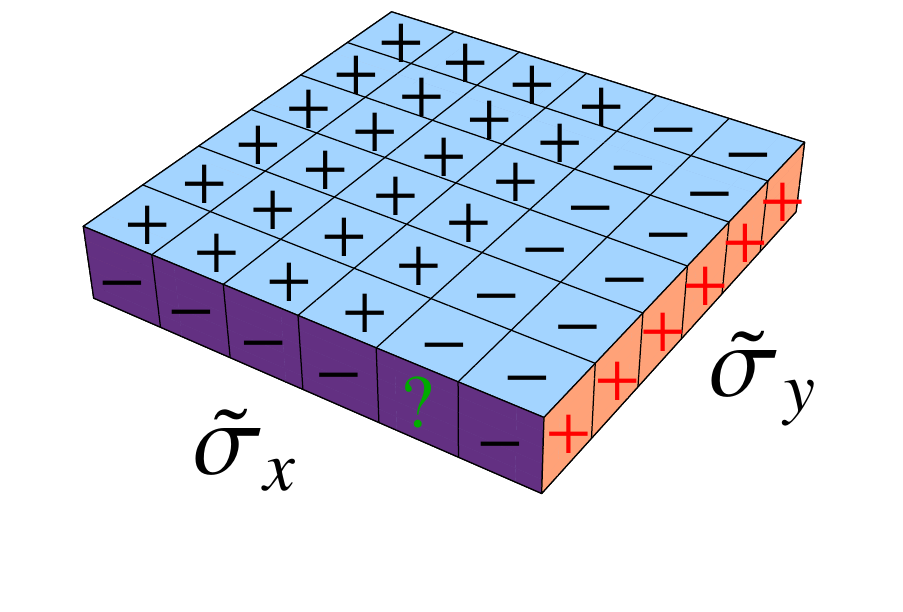}}
\subfigure[]{\includegraphics[width=0.45\columnwidth]{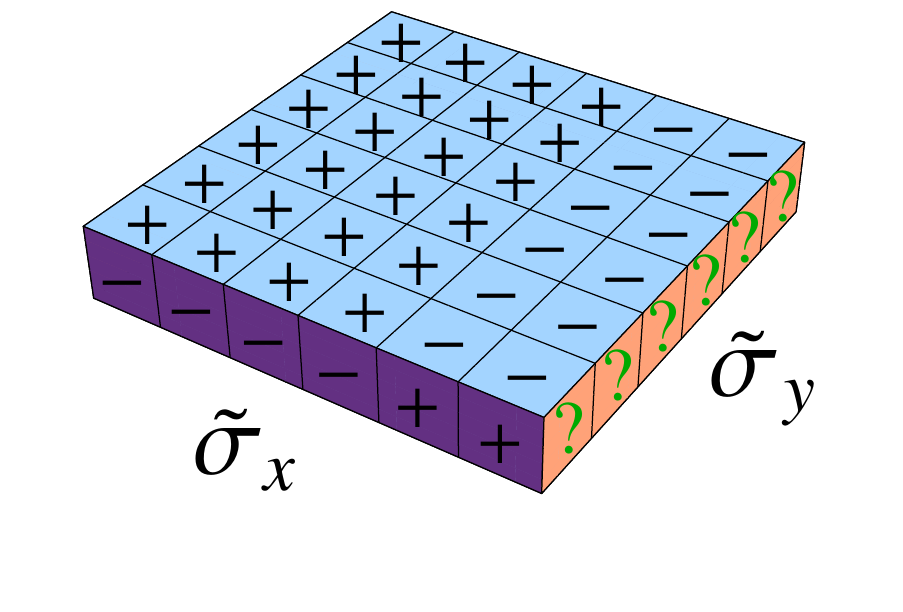}}\\
\caption{(a) A texture of type $C(4,2)$. (b) If at least one of the leftmost $p_{+}=4$ $\tilde{\sigma}_{x}$ spins is $+1$ (black), all the (red) $\tilde{\sigma}_{y}$ spins are $-1$. Subsequently the (purple) rightmost $p_{-}=2$ $\tilde{\sigma}_{y}$ spins are $+1$. (c) If all the leftmost $p_{+}=4$ $\tilde{\sigma}_{x}$ spins are $-1$ and at least one of the rightmost $p_{-}=2$ $\tilde{\sigma}_{x}$ spins (black) are $-1$, then all the (red) $\tilde{\sigma}_{y}$ spins are $+1$. (d) If all the leftmost $p_{+}=4$ $\tilde{\sigma}_{x}$ spins and all the rightmost $p_{-}=2$ $\tilde{\sigma}_{x}$ spins are $+1$, then there are no restrictions on the (green) $\tilde{\sigma}_{y}$ spins.}
\label{c1ex}
\end{figure}

In this case all the columns or rows are full, and we find that
\begin{align}
&Z^{C(p,L_{x}-p)}_{Q}(L_{x},L_{y})=\left(\begin{array}{c}L_{x}\\p\end{array}\right)\delta_{Q,2^{p}+2^{L_{x}-p}+2^{L_{y}}-2} ,\nonumber\\
&Z^{R(p,L_{y}-p)}_{Q}(L_{x},L_{y})=\left(\begin{array}{c}L_{y}\\p\end{array}\right)\delta_{Q,2^{p}+2^{L_{y}-p}+2^{L_{x}}-2} ,
\end{align}
see Fig. \ref{c1ex}.

\subsubsection{Types $C(p_{+},p_{-})$ and $R(p_{+},p_{-})$}
\begin{figure}
\subfigure[]{\includegraphics[width=0.45\columnwidth]{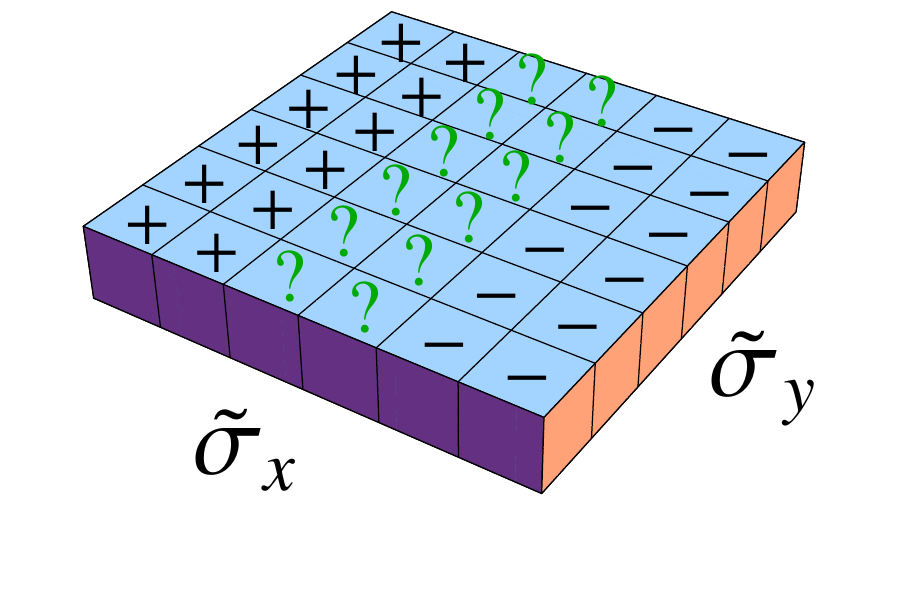}}
\subfigure[]{\includegraphics[width=0.45\columnwidth]{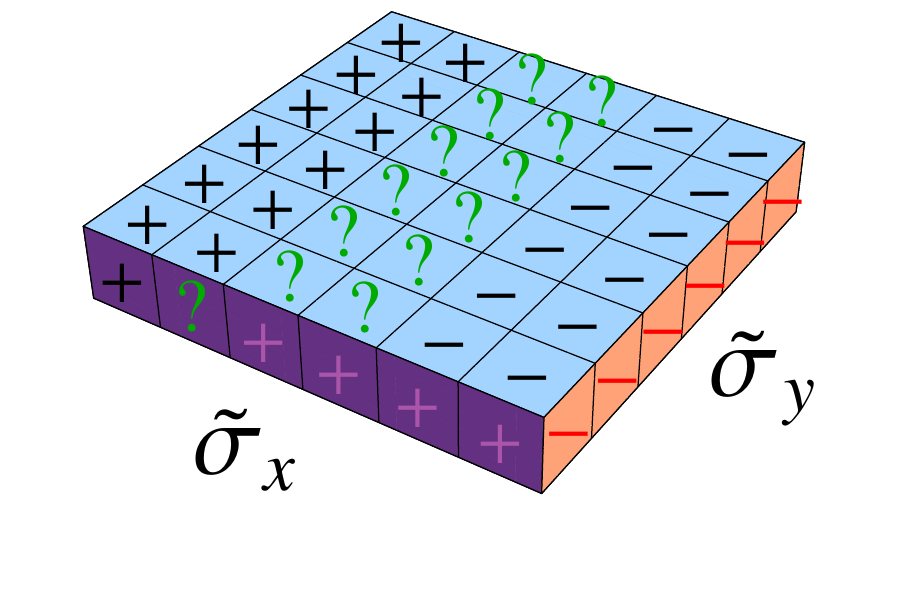}}\\
\subfigure[]{\includegraphics[width=0.45\columnwidth]{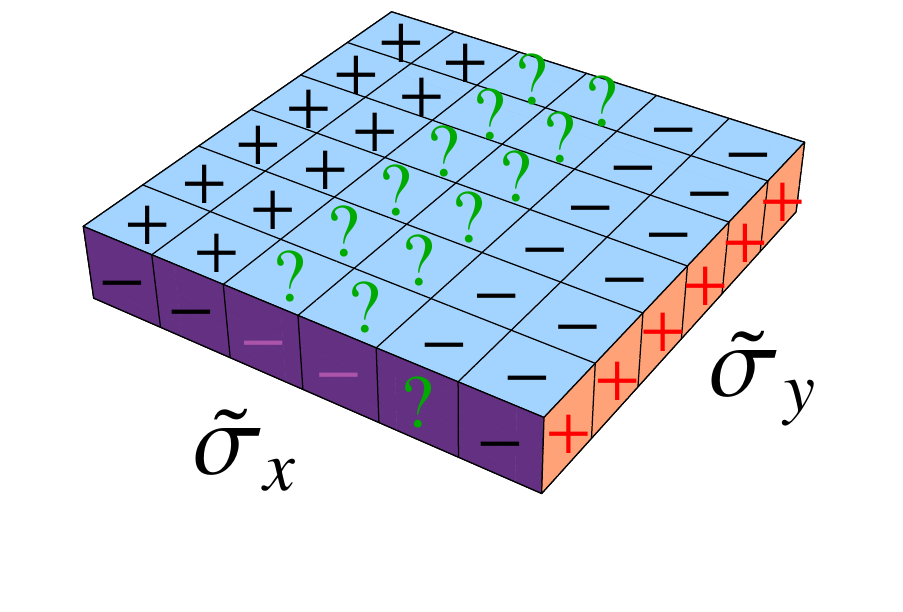}}
\subfigure[]{\includegraphics[width=0.45\columnwidth]{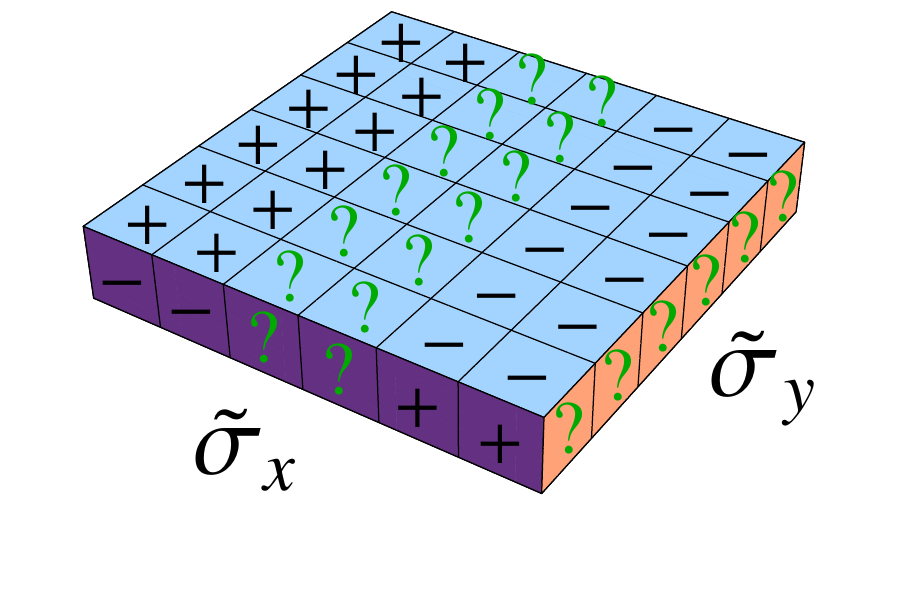}}\\
\caption{(a) A texture of type $C(2,2)$. Each of the two middle columns contains at least one $+1$ and one $-1$ spin. (b-d) A breakdown of all the possibilities for the $\tilde{\sigma}_{x}$ spins (in black), the constraints on the $\tilde{\sigma}_{y}$ spins (in red), and the subsequent constraints on the remaining $\tilde{\sigma}_{x}$ spins (in purple). Spins marked with a green question mark are free.}
\label{c2ex}
\end{figure}

We may assume without loss of generality that we consider a texture of type $C_{p_{+},p_{-}}$, that the $p_{+}$ full $+1$ columns are the leftmost columns, and that the $p_{-}$ full $-1$ columns are the rightmost columns. In the middle there is a block of size $(L_{x}-p_{+}-p_{-})\times L_{y}$ of type $0,R_{\pm}(k)$ or $R(k_{+},k_{-})$, such that in each of its columns there is at least one $+1$ spin and at least one $-1$ spin. See Fig. \ref{c2ex}a.

If at least one of the leftmost $p_{+}$ $\tilde{\sigma}_{x}$ spins is $+1$, then $\tilde{\sigma}_{y}=-1$, and subsequently all the rightmost $p_{-}$ $\tilde{\sigma}_{x}$ spins are $+1$. Since each of the middle columns has at least one $-1$ spin, the middle $L_{x}-p_{+}-p_{-}$ $\tilde{\sigma}_{x}$ spins are $+1$. See Fig. \ref{c2ex}b.

If all the leftmost $p_{+}$ $\tilde{\sigma}_{x}$ spins are $-1$, and at least one of the rightmost $p_{-}$ $\tilde{\sigma}_{x}$ spins is $-1$, then $\tilde{\sigma}_{y}=+1$. Since each of the middle columns has at least one $+1$ spin, the middle $L_{x}-p_{+}-p_{-}$ $\tilde{\sigma}_{x}$ spins are $-1$. See Fig. \ref{c2ex}c.

If all the leftmost $p_{+}$ $\tilde{\sigma}_{x}$ spins are $-1$, and all the rightmost $p_{-}$ $\tilde{\sigma}_{x}$ spins are $+1$, the only restriction comes from the middle columns. See Fig. \ref{c2ex}d.

Therefore, the recursion relations are
\begin{align}
&Z^{C(p_{+},p_{-})}_{Q}(L_{x},L_{y})=\nonumber\\
&=\left(\begin{array}{c}L_{x}\\p_{+}\end{array}\right)\left(\begin{array}{c}L_{x}-p_{+}\\p_{-}\end{array}\right)\delta_{Q,2^{p_{+}}+2^{p_{-}}-2+q}\nonumber\\
&\Biggl[\sum^{L_{y}-1}_{k_{+}=1}\sum^{L_{y}-k_{+}}_{k_{-}=1}Z^{R(k_{+},k_{-})}_{q}(L_{x}-p_{+}-p_{-},L_{y})+\nonumber\\
&\left.+\sum_{s=\pm1}\sum^{L_{y}-1}_{k=1}Z^{R_{s}(k)}_{q}(L_{x}-p_{+}-p_{-},L_{y})+\right.\nonumber\\
&+Z^{0}_{q}(L_{x}-p_{+}-p_{-},L_{y})\Biggr] ,\nonumber\\
&Z^{R(p_{+},p_{-})}_{Q}(L_{x},L_{y})=\nonumber\\
&=\left(\begin{array}{c}L_{y}\\p_{+}\end{array}\right)\left(\begin{array}{c}L_{y}-p_{+}\\p_{-}\end{array}\right)\delta_{Q,2^{p_{+}}+2^{p_{-}}-2+q}\nonumber\\
&\Biggl[\sum^{L_{x}-1}_{k_{+}=1}\sum^{L_{x}-k_{+}}_{k_{-}=1}Z^{C(k_{+},k_{-})}_{q}(L_{x},L_{y}-p_{+}-p_{-})+\nonumber\\
&\left.+\sum_{s=\pm1}\sum^{L_{x}-1}_{k=1}Z^{C_{s}(k)}_{q}(L_{x},L_{y}-p_{+}-p_{-})+\right.\nonumber\\
&+Z^{0}_{q}(L_{x},L_{y}-p_{+}-p_{-})\Biggr] .
\end{align}

\subsubsection{Types $C_{\pm}(p)$ and $R_{\pm}(p)$}
\begin{figure}
\subfigure[]{\includegraphics[width=0.45\columnwidth]{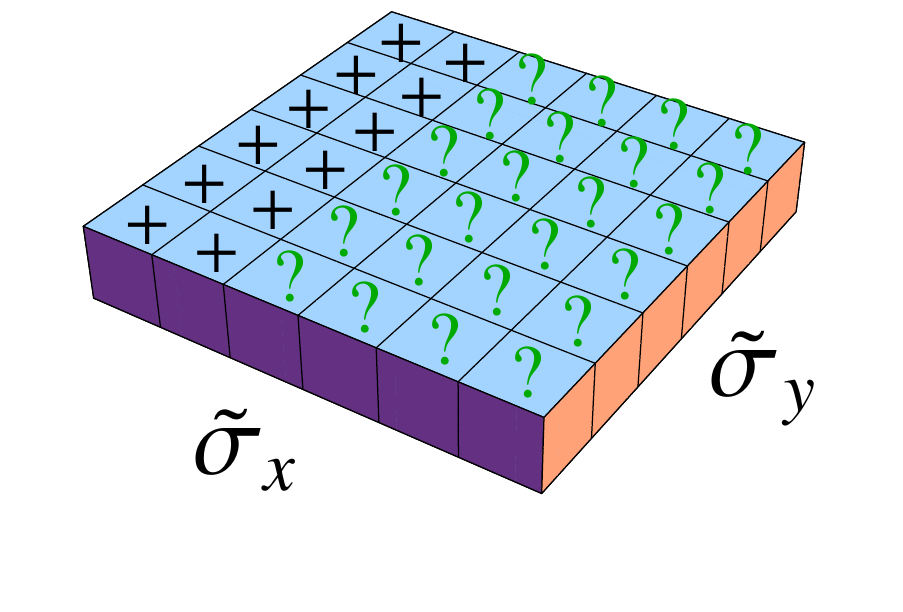}}\\
\subfigure[]{\includegraphics[width=0.45\columnwidth]{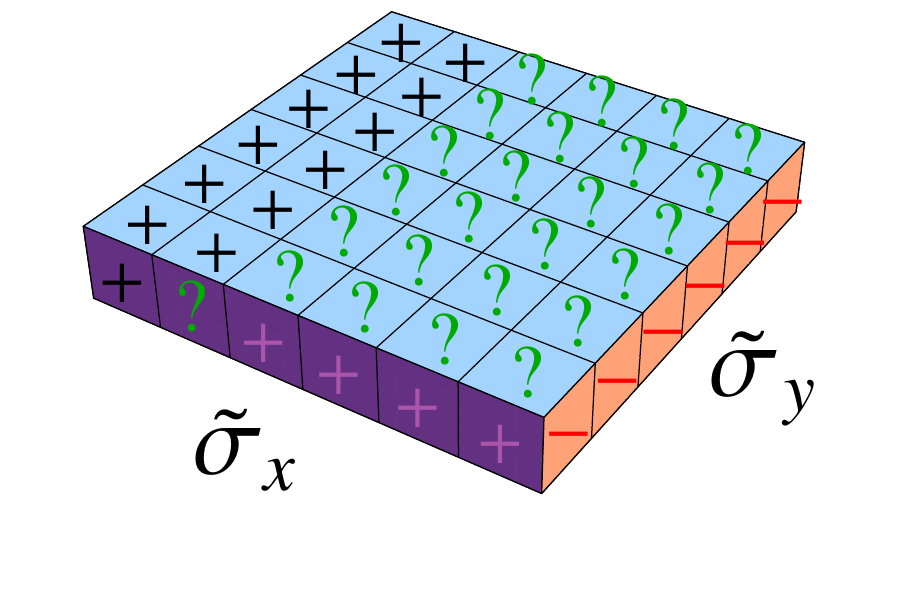}}
\subfigure[]{\includegraphics[width=0.45\columnwidth]{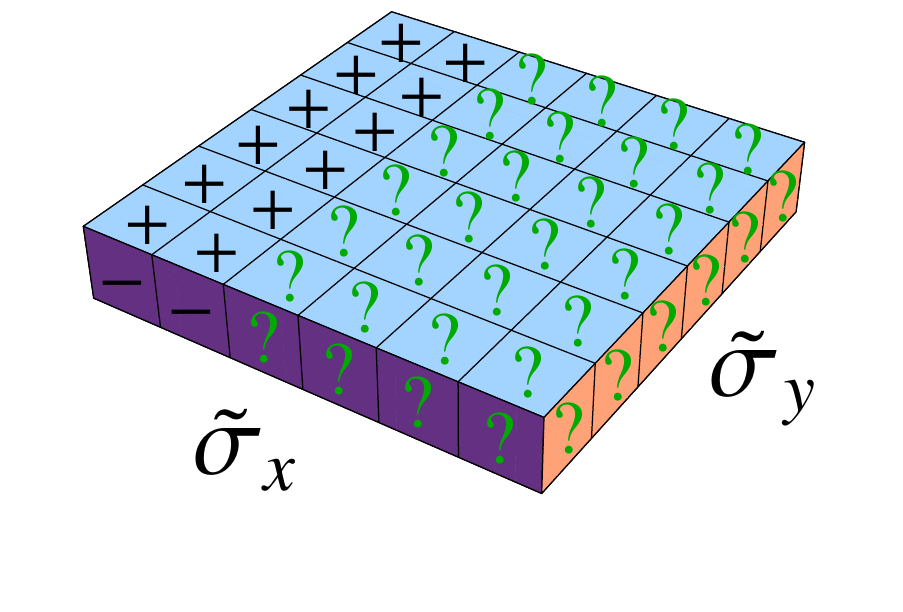}}
\caption{(a) A texture of type $C_{+}(2)$. Each of rightmost $L_{x}-p=4$ columns contains at least one $+1$ and one $-1$ spin. (b-c) A breakdown of all the possibilities for the $\tilde{\sigma}_{x}$ spins (in black), the constraints on the $\tilde{\sigma}_{y}$ spins (in red), and the subsequent constraints on the remaining $\tilde{\sigma}_{x}$ spins (in purple). Spins marked with a green question mark are free.}
\label{cpex}
\end{figure}

We assume without loss of generality that the texture is of type $C_{+}(p)$, and that the $p$ full $+1$ columns are the leftmost columns. The block of size $(L_{x}-p)\times L_{y}$ is either of type $0$ or $R_{-}(q)$, such that each columns contains at least one $-1$ spin, see Fig. \ref{cpex}a.

If at least one of the leftmost $p$ $\tilde{\sigma}_{x}$ spins is $+1$, then $\tilde{\sigma}_{y}=-1$. Since each of the rightmost $L_{x}-p$ columns contains at least one $-1$ spin, the rightmost $L_{x}-p$ $\tilde{\sigma}_{x}$ spins are $+1$. See Fig. \ref{cpex}b.

If all the leftmost $p$ $\tilde{\sigma}_{x}$ spins are $-1$, then the only restriction comes from the block, see Fig. \ref{cpex}c. Hence,
\begin{align}
&Z^{C_{\pm}(p)}_{Q}(L_{x},L_{y})=\left(\begin{array}{c}L_{x}\\p\end{array}\right)\delta_{Q,2^{p}-1+q}\nonumber\\
&\left[\sum^{L_{y}-1}_{k=1}Z^{R_{\mp}(k)}_{q}(L_{x}-p,L_{y})+Z^{0}_{q}(L_{x}-p,L_{y})\right] ,\nonumber\\
&Z^{R_{\pm}(p)}_{Q}(L_{x},L_{y})=\left(\begin{array}{c}L_{y}\\p\end{array}\right)\delta_{Q,2^{p}-1+q}\nonumber\\
&\left[\sum^{L_{x}-1}_{k=1}Z^{C_{\mp}(k)}_{q}(L_{x},L_{y}-p)+Z^{0}_{q}(L_{x},L_{y}-p)\right] .
\end{align}

\subsubsection{Type $0$}
\begin{figure}
\subfigure[]{\includegraphics[width=\columnwidth]{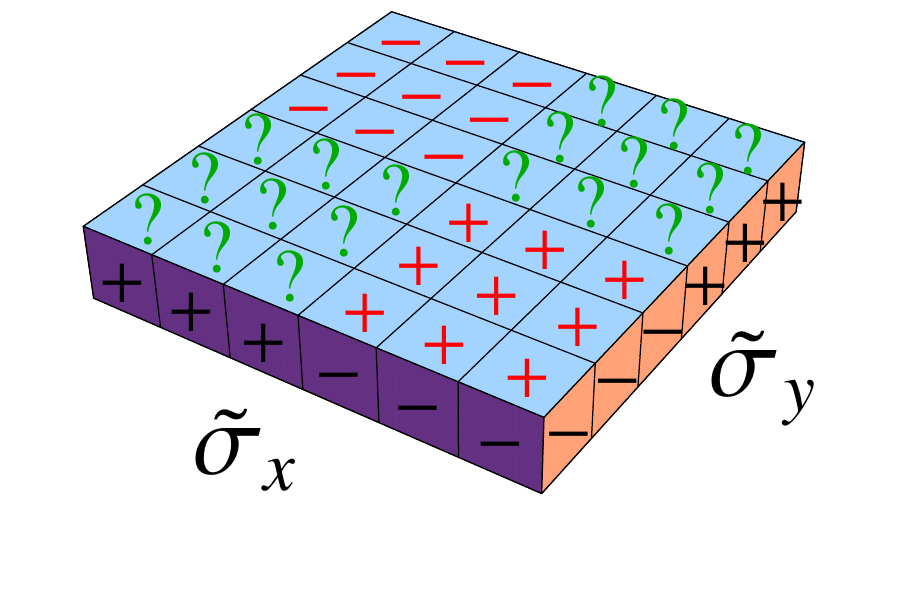}}
\caption{A texture of type $0$ which is part of a configuration with $p_{x}=3$ of its $\tilde{\sigma}_{x}$ spins equal to $+1$ and $p_{y}=3$ of its $\tilde{\sigma}_{y}$ spins equal to $+1$. The TL $3\times 3$ quadrant in the $\tilde{\sigma}_{z}$ texture is all $-1$, and the BR $3\times3$ quadrant is all $+1$.}
\label{z0figa}
\end{figure}

We assume that the $\tilde{\sigma}_{z}$ texture has at least one non-trivial solution with $1\leq p_{x}\leq L_{x}-1$ of the $\tilde{\sigma}_{x}$ spins and $1\leq p_{y}\leq L_{y}-1$ of the $\tilde{\sigma}_{y}$ spins in the $+1$ state. Without loss of generality, we assume that the $p_{x}$ $+1$ $\tilde{\sigma}_{x}$ spins are the leftmost spins and the $p_{y}$ $+1$ $\tilde{\sigma}_{y}$ spins are the topmost spins. Thus, the $\tilde{\sigma}_{z}$ texture is divided into four quadrants. All the spins in the $p_{x}\times p_{y}$ top-left quadrant are $-1$, all the spins in the $(L_{x}-p_{x})\times(L_{y}-p_{y})$ bottom-right quadrant are $+1$, while the spins in $(L_{x}-p_{x})\times p_{y}$ the top-right (TR) quadrant and the $p_{x}\times(L_{y}-p_{y})$ bottom-left (BL) quadrant are free. However, since the $L_{x}\times L_{y}$ $\tilde{\sigma}_{z}$ texture is of type $0$, we find that the TR quadrant is of type $0,C_{-}(k)$ or $R_{+}(k)$ and the BL quadrant is of type $0,C_{+}(k)$ or $R_{-}(k)$. In any case, in the TR quadrant all the columns contain at least one $-1$ spin and all the rows contain at least one $+1$ spin, and in the BL quadrant all the columns contain at least one $+1$ spin and all the rows contain at least one $-1$ spin. See Fig. \ref{z0figa}.

\begin{figure}
\subfigure[]{\includegraphics[width=0.45\columnwidth]{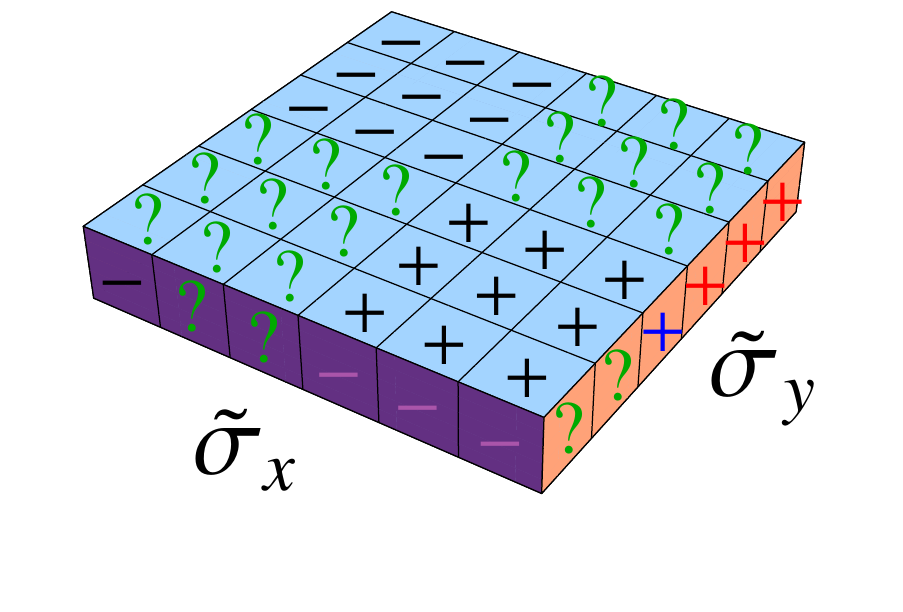}}
\subfigure[]{\includegraphics[width=0.45\columnwidth]{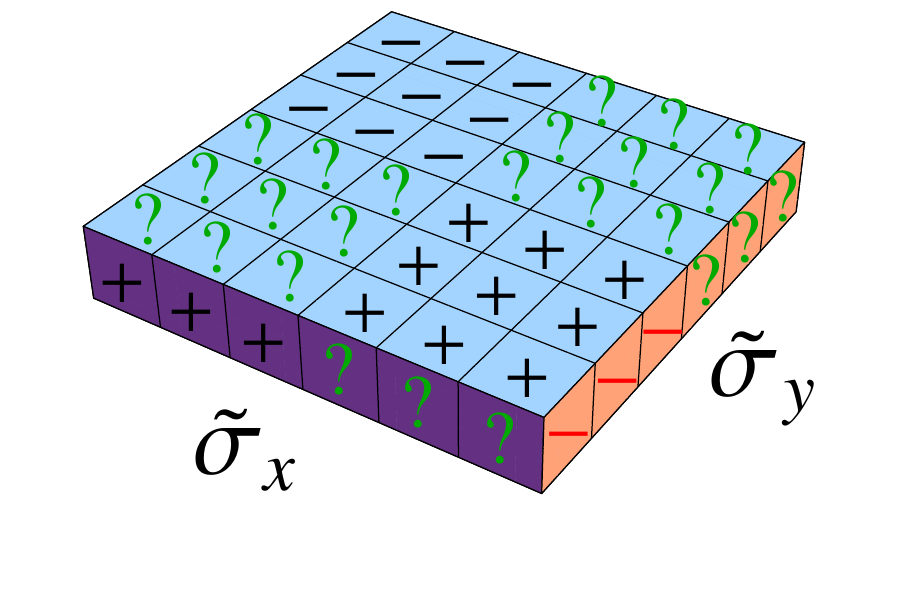}}
\caption{The BL quadrant is of type $0$. (a) If at least one of the leftmost $p_{x}$ $\tilde{\sigma}_{x}$ spins is $-1$ (black), the $p_{y}=3$ topmost $\tilde{\sigma}_{y}$ spins are $-1$ (red). Since the BL quadrant is of type $0$, each column in it contains at least one $-1$ spin and thus at least one of the bottom $\tilde{\sigma}_{y}$ spins is $+1$ (blue). From the intersection of the blue $\tilde{\sigma}_{y}$ spin and the BR quadrant, the rightmost (purple) $\tilde{\sigma}_{x}$ spins are $-1$. (b) If all the leftmost $\tilde{\sigma}_{x}$ spins are $+1$, then all the bottom $\tilde{\sigma}_{y}$ spins are $-1$, since each row in the BR quadrant contains a $+1$ spin.}
\label{z0figb}
\end{figure}

Now consider the $q_{1}$ solutions of the BL quadrant and the $q_{2}$ solutions of the TR quadrant. Each solution has at least one $\tilde{\sigma}_{x}$ spin equal to $+1$ and one $\tilde{\sigma}_{y}$ spin equal to $-1$, or at least one $\tilde{\sigma}_{x}$ spin equal to $-1$ and one $\tilde{\sigma}_{y}$ spin equal to $+1$, otherwise $\tilde{\sigma}^{x}=\tilde{\sigma}^{y}=\pm1$, which is possible only if the blocks are of type $CR_{\mp}$, and they are not. Therefore we should consider the following cases:

\paragraph{BL or TR of type $0$}
In this case, each column and row in BL and TR contains at least one $+1$ spin and one $-1$ spin. If one of the leftmost $p_{x}$ $\tilde{\sigma}_{x}$ spins is $-1$, then the topmost $p_{y}$ $\tilde{\sigma}_{y}$ spins are $+1$ and at least one of the bottom $L_{y}-p_{y}$ $\tilde{\sigma}_{y}$ spins is $+1$. Subsequently, the rightmost $L_{x}-p_{x}$ $\tilde{\sigma}_{x}$ spins are $-1$. See Fig. \ref{z0figb}.

If all the leftmost $p_{x}$ $\tilde{\sigma}_{x}$ spins are $+1$, then the bottom $L_{y}-p_{y}$ $\tilde{\sigma}_{y}$ spins are $-1$ and the only remaining restriction comes from TR. See Fig. \ref{z0figb}a.

A similar reasoning can be done starting from the TR quadrant, and thus if at least one of BL or TR is of type $0$, we find that $Q=q_{1}+q_{2}-1$. See Fig. \ref{z0figb}b.

\begin{figure}
\subfigure[]{\includegraphics[width=0.45\columnwidth]{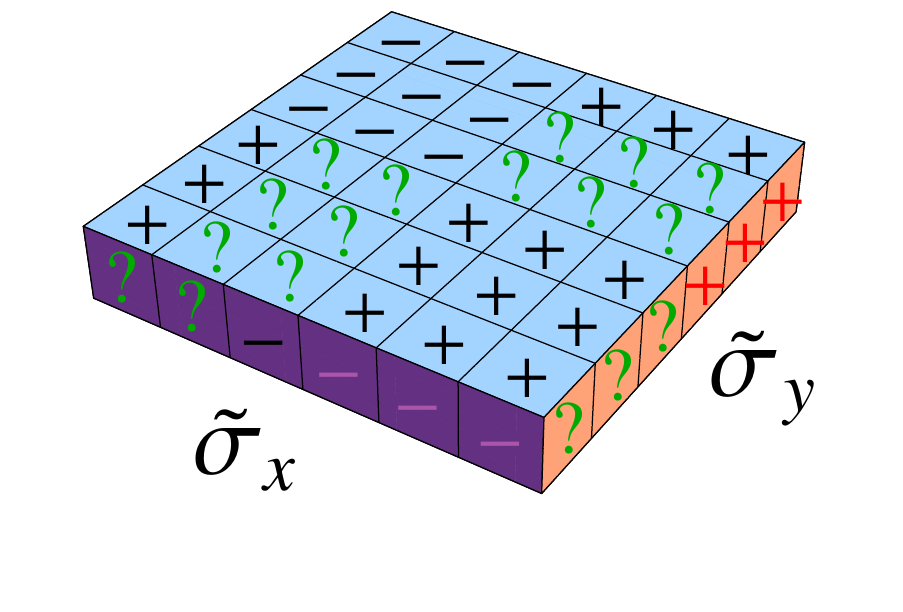}}
\subfigure[]{\includegraphics[width=0.45\columnwidth]{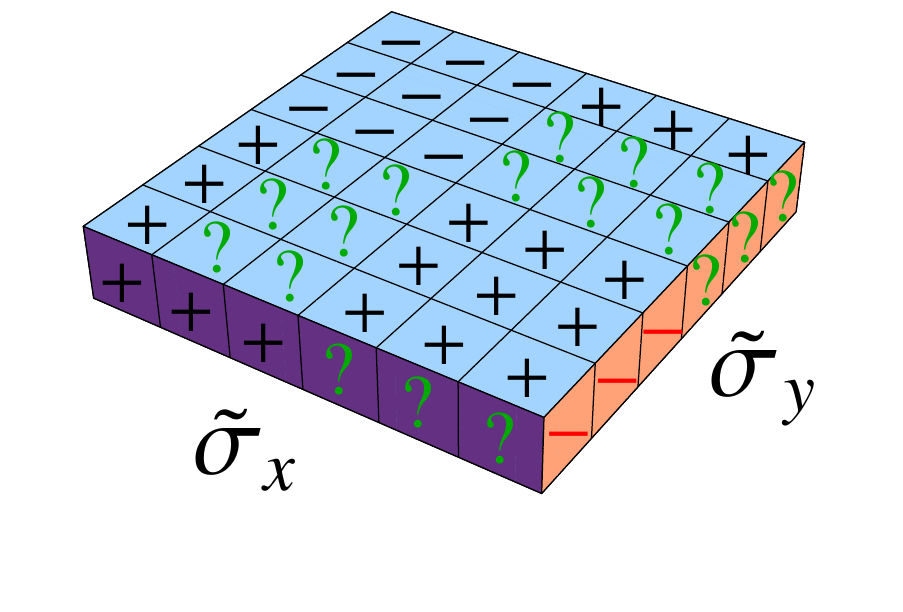}}
\caption{The BL quadrant is of type $C_{+}(1)$ and the TR quadrant is of type $R_{+}(1)$. (a) If at least one of the middle left $p_{x}-k_{1}=2$ $\tilde{\sigma}_{x}$ spins is $-1$ (black), the topmost (red) $\tilde{\sigma}_{y}$ spins are $+1$. From the intersection of the (red) $\tilde{\sigma}_{y}$ spins and the top row in the TR quadrant, the rightmost (purple) $\tilde{\sigma}_{x}$ spins are $-1$. (b) If all the leftmost $\tilde{\sigma}_{x}$ spins (black) are $+1$, then from their intersection with the leftmost column in the BL quadrant, the bottom $\tilde{\sigma}_{y}$ spins (red) are $-1$. The other $\tilde{\sigma}_{x}$ and $\tilde{\sigma}_{y}$ spins are determined by the TR quadrant.}
\label{z0figc}
\end{figure}

\paragraph{BL of type $C_{+}(k_{1})$ and TR of type $R_{+}(k_{2})$, or BL of type $R_{-}(k_{1})$ and TR of type $C_{-}(k_{2})$}
We assume without loss of generality that BL is of type $C_{+}(k_{1})$, and that the $k_{1}$ $+1$ columns are the leftmost. See Fig. \ref{z0figc}.

If at least one of the leftmost $p_{x}$ $\tilde{\sigma}_{x}$ spins is $-1$, then the topmost $p_{y}$ $\tilde{\sigma}_{y}$ spins are $+1$. Subsequently, the rightmost $L_{x}-p_{x}$ $\tilde{\sigma}_{x}$ spins are $-1$. See Fig. \ref{z0figc}a.

If all the leftmost $p_{x}$ $\tilde{\sigma}_{x}$ spins are $+1$, then the bottom $L_{y}-p_{y}$ $\tilde{\sigma}_{y}$ spins are $-1$, and the only further restriction comes from TR. See Fig. \ref{z0figc}b.
A similar reasoning can be done starting from the TR quadrant, and thus $Q=q_{1}+q_{2}-1$.

\begin{figure}
\subfigure[]{\includegraphics[width=0.3\columnwidth]{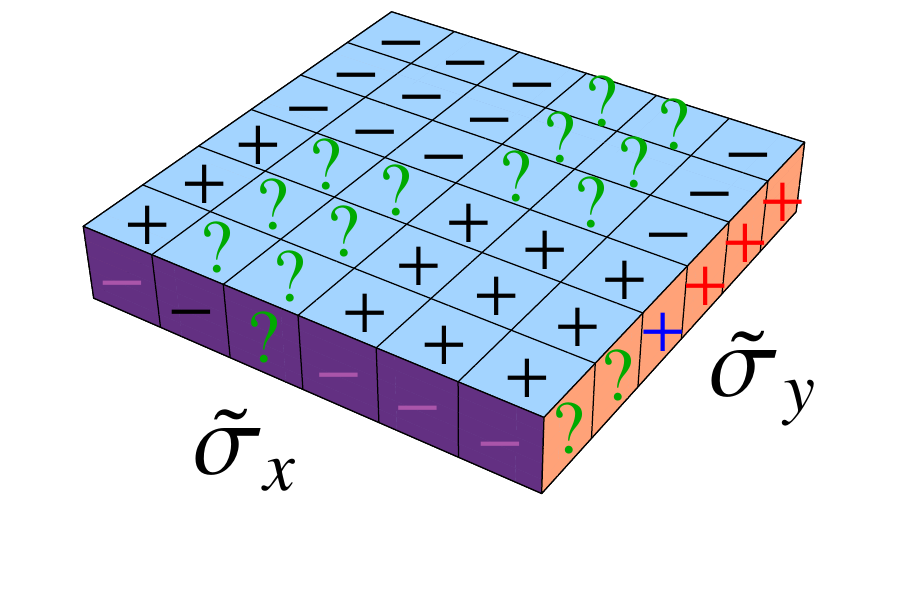}}
\subfigure[]{\includegraphics[width=0.3\columnwidth]{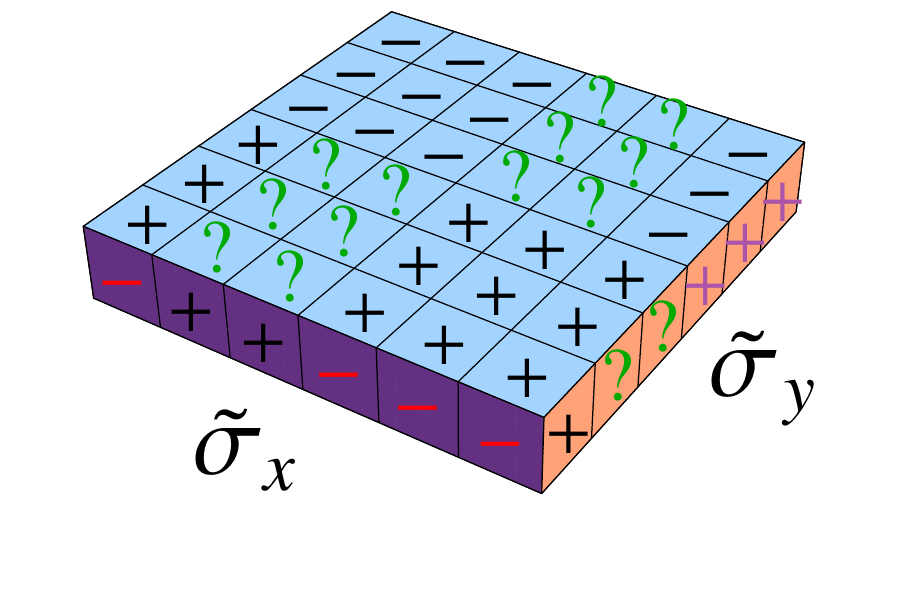}}
\subfigure[]{\includegraphics[width=0.3\columnwidth]{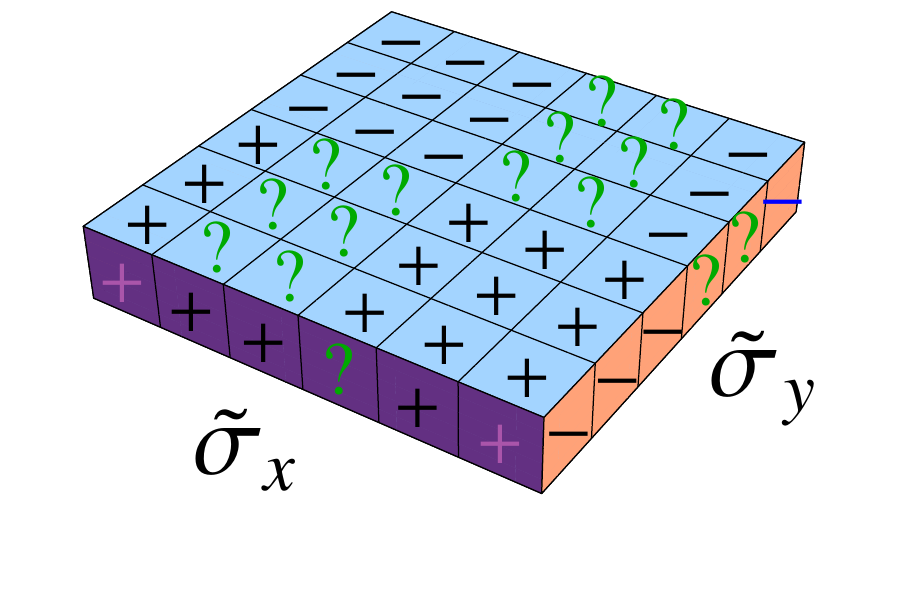}}\\
\subfigure[]{\includegraphics[width=0.3\columnwidth]{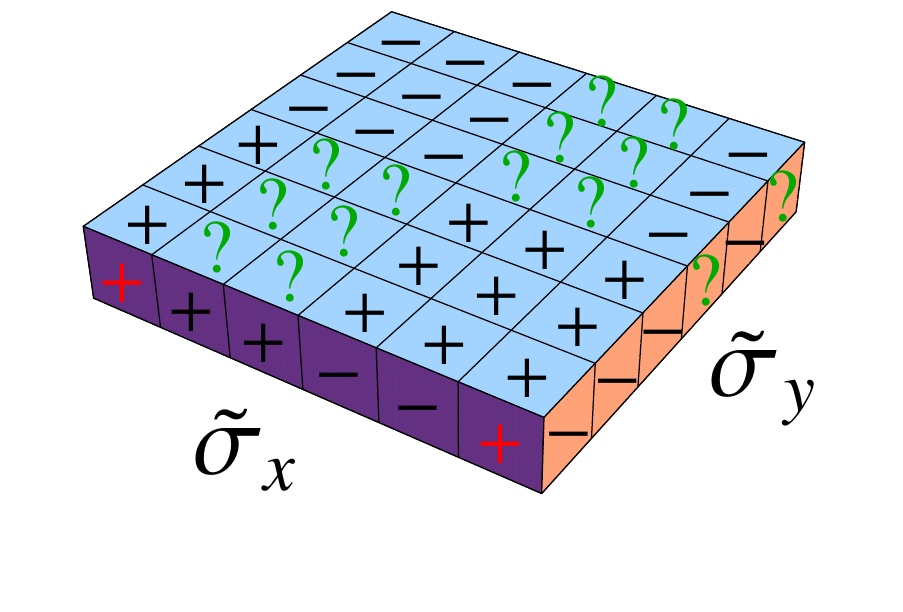}}
\subfigure[]{\includegraphics[width=0.3\columnwidth]{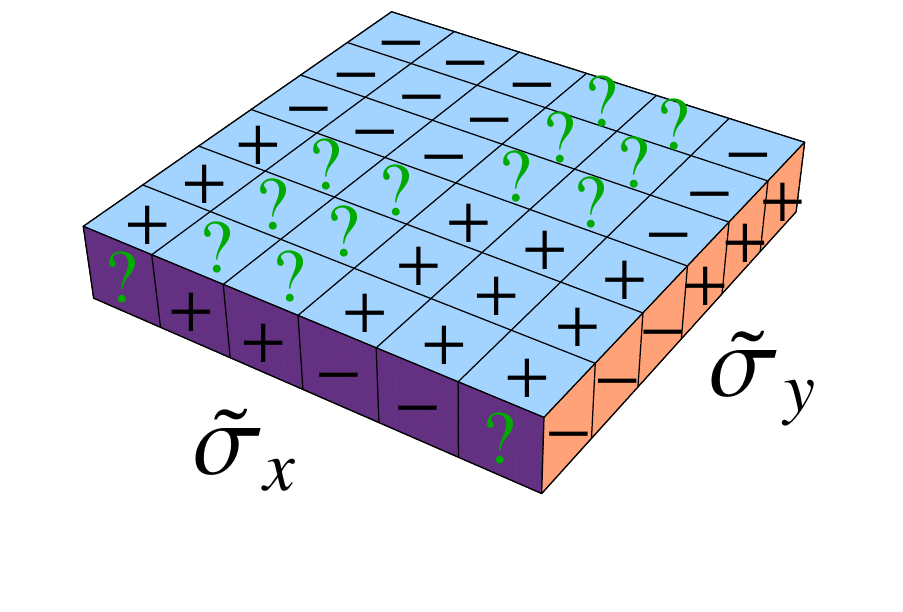}}
\caption{The BL quadrant is of type $C_{+}(1)$ and the TR quadrant is of type $C_{-}(1)$. (a) If at least one of the left-middle $p_{x}-k_{1}=2$ $\tilde{\sigma}_{x}$ spins is $-1$ (black), the topmost $\tilde{\sigma}_{y}$ spins (red) are $+1$. Because each row in the $BL$ quadrant, except the leftmost $k_{1}$ columns, contains at least one $-1$ spin, at least one of the bottom $\tilde{\sigma}_{y}$ spins (blue) is $+1$. From the intersection of the blue $\tilde{\sigma}_{y}$ spin with the $\tilde{\sigma}_{z}$ spins, the rightmost and leftmost $\tilde{\sigma}_{x}$ spins are $-1$ (purple). (a) The black spins are set as shown. The remaining (red) $\tilde{\sigma}_{x}$ spins are therefore $-1$, and subsequently the top (purple) $\tilde{\sigma}_{y}$ spins are $+1$. (c) The black spins are set as shown. Since each column in the TR quadrant contains at least one $+1$ spin, at least one of the topmost $\tilde{\sigma}_{y}$ spins is $-1$ (blue). From the intersection of the blue spin, the leftmost and rightmost (purple) $\tilde{\sigma}_{x}$ spins are $+1$. (d) The black spins are set as shown, and they constrain the red spins as shown. (e) The black spins are set as shown. The remaining green $k_{1}+k_{2}=2$ $\tilde{\sigma}_{x}$ spins are free.}
\label{z0figd}
\end{figure}

\paragraph{BL of type $C_{+}(k_{1})$ and TR of type $C_{-}(k_{2})$, or BL of type $R_{-}(k_{1})$ and TR of type $R_{+}(k_{2})$.}
We assume without loss of generality that BL is of type $C_{+}(k_{1})$, that the $k_{1}$ $+1$ columns are the leftmost, and that the $k_{2}$ $-1$ columns are the rightmost. See Fig. \ref{z0figd}.

If at least one of the left-middle $p_{x}-k_{1}$ $\tilde{\sigma}_{x}$ spins is $-1$, then the topmost $p_{y}$ $\tilde{\sigma}_{y}$ spins and at least one of the bottom $L_{y}-p_{y}$ $\tilde{\sigma}_{y}$ spins are $+1$, and thus the rightmost $L_{x}-p_{x}$ $\tilde{\sigma}_{x}$ spins are $-1$. See Fig. \ref{z0figd}a.

If all the left-middle $p_{x}-k_{1}$ $\tilde{\sigma}_{x}$ spins are $+1$, and at least one of the bottom $L_{y}-p_{y}$ $\tilde{\sigma}_{y}$ spins is $+1$, then the leftmost $k_{1}$ and the rightmost $L_{x}-p_{x}$ $\tilde{\sigma}_{x}$ spins are $-1$, and subsequently all the topmost $p_{y}$ $\tilde{\sigma}_{y}$ spins are $+1$. See Fig. \ref{z0figd}b. The last two cases give a total of $q_{1}-2^{k_{1}}$ solutions. 

If all the left-middle $p_{x}-k_{1}$ $\tilde{\sigma}_{x}$ spins are $+1$, all the bottom $L_{y}-p_{y}$ $\tilde{\sigma}_{y}$ spins are $-1$, and at least one of the right-middle $L_{x}-p_{x}-k_{2}$ $\tilde{\sigma}_{x}$ spins is $+1$, then at least one of the top $p_{y}$ $\tilde{\sigma}_{y}$ spins is $-1$. Subsequently, all the leftmost $k_{1}$ and the rightmost $k_{2}$ $\tilde{\sigma}_{x}$ spins are $+1$. See Fig. \ref{z0figd}c.

If all the left-middle $p_{x}-k_{1}$ $\tilde{\sigma}_{x}$ spins are $+1$, all the bottom $L_{y}-p_{y}$ $\tilde{\sigma}_{y}$ spins are $-1$, all the right-middle $L_{x}-p_{x}-k_{2}$ $\tilde{\sigma}_{x}$ spins are $-1$, and at least one of the top $p_{y}$ $\tilde{\sigma}_{y}$ spins is $-1$, then the leftmost $k_{1}$ and the rightmost $k_{2}$ $\tilde{\sigma}_{x}$ spins are $+1$. See Fig. \ref{z0figd}d. These last two cases give a total of $q_{2}-2^{k_{2}}$ solutions.

If all the left-middle $p_{x}-k_{1}$ $\tilde{\sigma}_{x}$ spins are $+1$, all the bottom $L_{y}-p_{y}$ $\tilde{\sigma}_{y}$ spins are $-1$, all the right-middle $L_{x}-p_{x}-k_{2}$ $\tilde{\sigma}_{x}$ spins are $-1$, and all the top $p_{y}$ $\tilde{\sigma}_{y}$ spins are $+1$, the remaining leftmost $k_{1}$ and rightmost $k_{2}$ $\tilde{\sigma}_{x}$ spins are free, which gives another $2^{k_{1}+k_{2}}$ solutions. See Fig. \ref{z0figd}e.

Combining all of the above, we obtain the following recursion relation for the type $0$ textures:
\begin{widetext}
\begin{align}
&Z^{0}_{Q}\left(L_{x},L_{y}\right)=\frac{1}{Q-2}\sum^{L_{x}-1}_{p_{x}=1}\sum^{L_{y}-1}_{p_{y}=1}\left(\begin{array}{c}L_{x}\\p_{x}\end{array}\right)\left(\begin{array}{c}L_{y}\\p_{y}\end{array}\right)\left\{\sum^{Q-1}_{q=2}Z^{0}_{q}(L_{x}-p_{x},p_{y})Z^{0}_{Q+1-q}(p_{x},L_{y}-p_{y})+\right.\nonumber\\
&\left.+Z^{0}_{Q+1-q}(L_{x}-p_{x},p_{y})\left[\sum^{p_{x}-1}_{k=1}Z^{C_{+}(k)}_{q}(p_{x},L_{y}-p_{y})+\sum^{L_{y}-p_{y}-1}_{k=1}Z^{R_{-}(k)}_{q}(p_{x},L_{y}-p_{y})\right]+\right.\nonumber\\
&\left.+Z^{0}_{Q+1-q}(p_{x},L_{y}-p_{y})\left[\sum^{L_{x}-p_{x}-1}_{k=1}Z^{C_{-}(k)}_{q}(L_{x}-p_{x},p_{y})+\sum^{p_{y}-1}_{k=1}Z^{R_{+}(k)}_{q}(L_{x}-p_{x},p_{y})\right]+\right.\nonumber\\
&\left.+\sum^{p_{x}-1}_{k_{1}=1}\sum^{p_{y}-1}_{k_{2}=1}Z^{C_{+}(k_{1})}_{q}(p_{x},L_{y}-p_{y})Z^{R_{+}(k_{2})}_{Q+1-q}(L_{x}-p_{x},p_{y})+\right.\nonumber\\
&\left.+\sum^{L_{y}-p_{y}-1}_{k_{1}=1}\sum^{L_{x}-p_{x}-1}_{k_{2}=1}Z^{R_{-}(k_{1})}_{q}(p_{x},L_{y}-p_{y})Z^{C_{-}(k_{2})}_{Q+1-q}(L_{x}-p_{x},p_{y})+\right.\nonumber\\
&\left.+\sum^{p_{x}-1}_{k_{1}=1}\sum^{L_{x}-p_{x}-1}_{k_{2}=1}Z^{C_{+}(k_{1})}_{q}(p_{x},L_{y}-p_{y})Z^{C_{-}(k_{2})}_{Q-q-2^{k_{1}+k_{2}}+2^{k_{1}}+2^{k_{2}}}(L_{x}-p_{x},p_{y})+\right.\nonumber\\
&\left.+\sum^{L_{y}-p_{y}-1}_{k_{1}=1}\sum^{p_{y}-1}_{k_{2}=1}Z^{R_{-}(k_{1})}_{q}(p_{x},L_{y}-p_{y})Z^{R_{+}(k_{2})}_{Q-q-2^{k_{1}+k_{2}}+2^{k_{1}}+2^{k_{2}}}(L_{x}-p_{x},p_{y})\right\} .
\end{align}
\end{widetext}
The $Q-2$ in the denominator appears because each solution was counted $Q-2$ times at each possible combination of $p_{x}$ and $p_{y}$.

\subsubsection{Final result}
\label{finalresult}
Combining all the above, including the symmetry relations, the recursion relations may be written as
\begin{widetext}
\begin{align}
&Z_{Q}(L_{x},L_{y})=Z^{0}_{Q}(L_{x},L_{y})+2Z^{C_{+}}_{Q}(L_{x},L_{y})+2Z^{C_{+}}_{Q}(L_{y},L_{x})+Z^{C}_{Q}(L_{x},L_{y})+Z^{C}_{Q}(L_{y},L_{x})+\nonumber\\
&+2Z^{CR_{+}}_{Q}(L_{x},L_{y}) ,\nonumber\\
&Z^{0}_{Q}\left(L_{x},L_{y}\right)=\frac{1}{Q-2}\sum^{L_{x}-1}_{p_{x}=1}\sum^{L_{y}-1}_{p_{y}=1}\sum^{Q-1}_{q=2}\left(\begin{array}{c}L_{x}\\p_{x}\end{array}\right)\left(\begin{array}{c}L_{y}\\p_{y}\end{array}\right)\nonumber\\
&\left\{Z^{0}_{Q+1-q}(L_{x}-p_{x},L_{y}-p_{y})\left[2Z^{C_{+}}_{q}(p_{x},p_{y})+2Z^{C_{+}}_{q}(p_{y},p_{x})+Z^{0}_{q}(p_{x},p_{y})\right]+\right.\nonumber\\
&\left.+2Z^{C_{+}}_{q}(p_{x},L_{y}-p_{y})Z^{C_{+}}_{Q+1-q}(p_{y},L_{x}-p_{x})+\right.\nonumber\\
&\left.+\sum^{p_{x}-1}_{k_{1}=1}\sum^{L_{x}-p_{x}-1}_{k_{2}=1}\left(\begin{array}{c}p_{x}\\k_{1}\end{array}\right)\left(\begin{array}{c}L_{x}-p_{x}\\k_{2}\end{array}\right)\left[Z^{C_{+}}_{q+1-2^{k_{1}}}(p_{y},p_{x}-k_{1})+Z^{0}_{q+1-2^{k_{1}}}(p_{x}-k_{1},p_{y})\right]\times\right.\nonumber\\
&\left.\times\left[Z^{C_{+}}_{Q-q+1-2^{k_{1}+k_{2}}+2^{k_{1}}}(L_{y}-p_{y},L_{x}-p_{x}-k_{2})+Z^{0}_{Q-q+1-2^{k_{1}+k_{2}}+2^{k_{1}}}(L_{x}-p_{x}-k_{2},L_{y}-p_{y})\right]+\right.\nonumber\\
&\left.+\sum^{p_{y}-1}_{k_{1}=1}\sum^{L_{y}-p_{y}-1}_{k_{2}=1}\left(\begin{array}{c}p_{y}\\k_{1}\end{array}\right)\left(\begin{array}{c}L_{y}-p_{y}\\k_{2}\end{array}\right)\left[Z^{C_{+}}_{q+1-2^{k_{1}}}(p_{x},p_{y}-k_{1})+Z^{0}_{q+1-2^{k_{1}}}(p_{y}-k_{1},p_{x})\right]\times\right.\nonumber\\
&\left.\times\left[Z^{C_{+}}_{Q-q+1-2^{k_{1}+k_{2}}+2^{k_{1}}}(L_{x}-p_{x},L_{y}-p_{y}-k_{2})+Z^{0}_{Q-q+1-2^{k_{1}+k_{2}}+2^{k_{1}}}(L_{y}-p_{y}-k_{2},L_{x}-p_{x})\right]\right\}\nonumber\\
&Z^{C_{+}}_{Q}(L_{x},L_{y})=\sum^{L_{x}-1}_{p=1}\left(\begin{array}{c}L_{x}\\p\end{array}\right)\delta_{Q,2^{p}-1+q}\left[Z^{C_{+}}_{q}(L_{y},L_{x}-p)+Z^{0}_{q}(L_{x}-p,L_{y})\right] ,\nonumber\\
&Z^{C}_{Q}(L_{x},L_{y})=\sum^{L_{x}-1}_{p_{+}=1}\sum^{L_{x}-p_{+}-1}_{p_{-}=1}\left(\begin{array}{c}L_{x}\\p_{+}\end{array}\right)\left(\begin{array}{c}L_{x}-p_{+}\\p_{-}\end{array}\right)\delta_{Q,2^{p_{+}}+2^{p_{-}}-2+q}\left[Z^{C}_{q}(L_{y},L_{x}-p_{+}-p_{-})+\right.\nonumber\\
&\left.+2Z^{C_{+}}_{q}(L_{y},L_{x}-p_{+}-p_{-})+Z^{0}_{q}(L_{x}-p_{+}-p_{-},L_{y})\right]+\sum^{L_{x}-1}_{p=1}\left(\begin{array}{c}L_{x}\\p\end{array}\right)\delta_{Q,2^{p}+2^{L_{x}-p}+2^{L_{y}}-2} ,\nonumber\\
&Z^{CR_{+}}_{Q}(L_{x},L_{y})=\sum^{L_{x}-1}_{p_{x}=1}\sum^{L_{y}-1}_{p_{y}=1}\left(\begin{array}{c}L_{x}\\p_{x}\end{array}\right)\left(\begin{array}{c}L_{y}\\p_{y}\end{array}\right)\delta_{Q,2^{p_{x}}+2^{p_{y}}-2+q}\left[Z^{CR_{+}}_{q}(L_{x}-p_{x},L_{y}-p_{y})+\right.\nonumber\\
&\left.+Z^{C_{+}}_{q}(L_{x}-p_{x},L_{y}-p_{y})+Z^{C_{+}}_{q}(L_{y}-p_{y},L_{x}-p_{x})+Z^{0}_{q}(L_{x}-p_{x},L_{y}-p_{y})\right]+\delta_{Q,2^{L_{x}}+2^{L_{y}}-1} ,
\end{align}
\end{widetext}
where we used
\begin{align}
&Z^{C}_{Q}(L_{x},L_{y})=\sum^{L_{x}-1}_{p_{+}=1}\sum^{L_{x}-p_{+}}_{p_{-}=1}Z^{C(p_{+},p_{-})}_{Q}(L_{x},L_{y}) ,\nonumber\\
&Z^{C_{+}}_{Q}(L_{x},L_{y})=\sum^{L_{x}-1}_{p=1}Z^{C_{+}(p)}_{Q}(L_{x},L_{y}) ,\nonumber\\
&Z^{CR_{+}}_{Q}(L_{x},L_{y})=\sum^{L_{x}}_{p_{x}=1}\sum^{L_{y}}_{p_{y}=1}Z^{CR_{+}(p_{x},p_{y})}_{Q}(L_{x},L_{y}) .
\end{align}
Note that in order to find $Z^{\beta}_{Q}(L_{x},L_{y})$ for $Q\geq3$, we don't need to know $Z^{0}_{2}(L_{x},L_{y})$, only $Z_{2}$ for smaller systems. Thus, $Z^{0}_{2}(L_{x},L_{y})$ is determined by
\begin{align}
Z^{0}_{2}(L_{x},L_{y})=2^{L_{x}L_{y}}-\sum_{Q\geq3}Z_{Q} .
\end{align}

\begin{figure}[b]
\includegraphics[width=0.49\columnwidth]{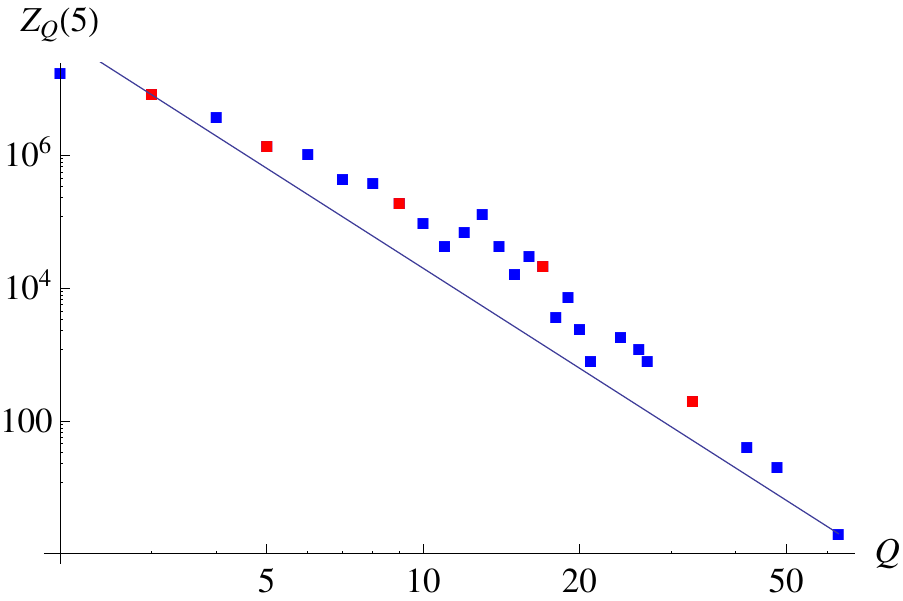}
\includegraphics[width=0.49\columnwidth]{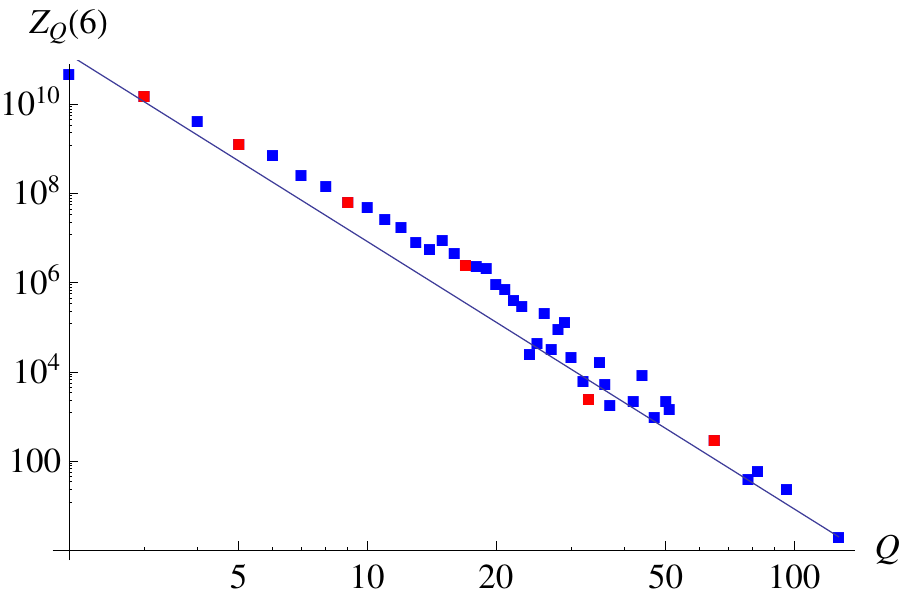}\\
\includegraphics[width=0.49\columnwidth]{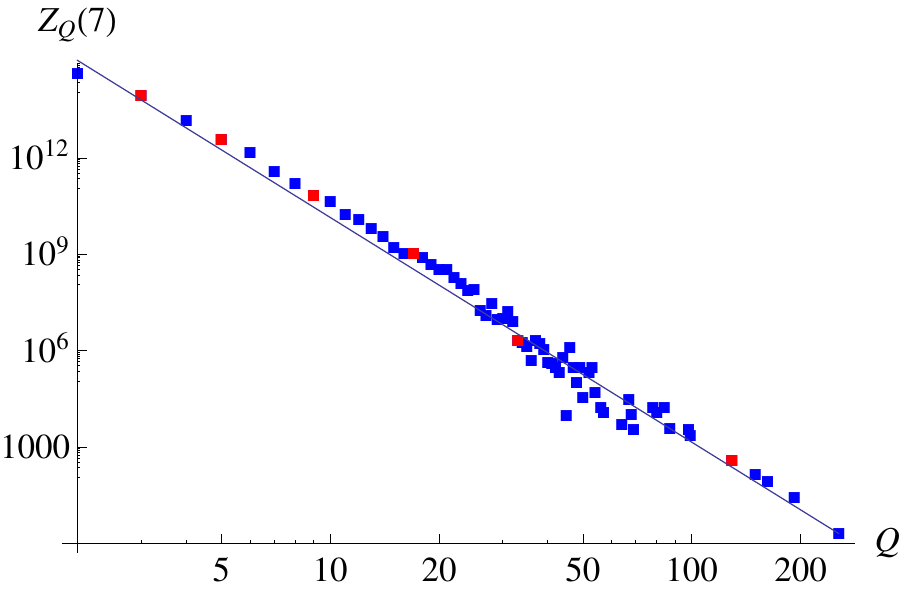}
\includegraphics[width=0.49\columnwidth]{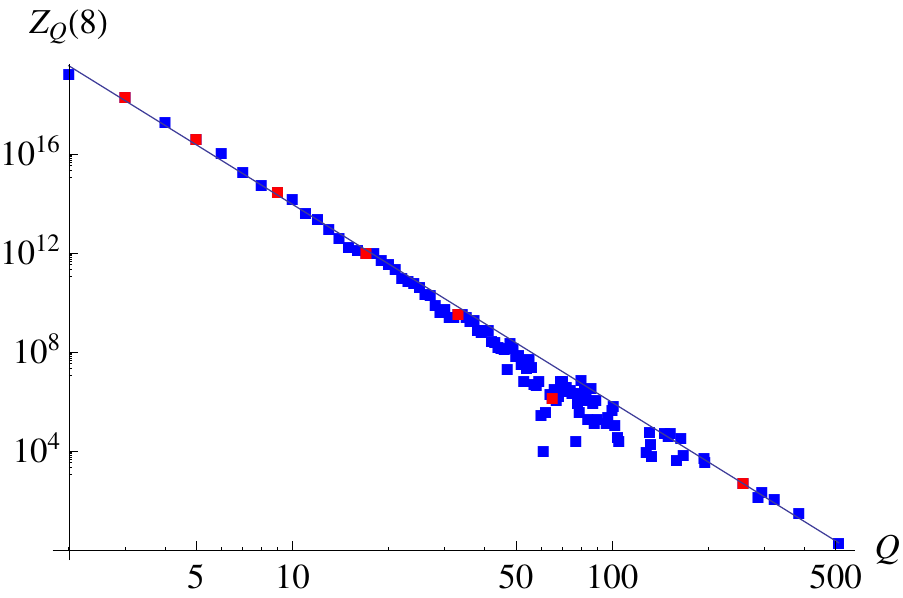}\\
\includegraphics[width=0.49\columnwidth]{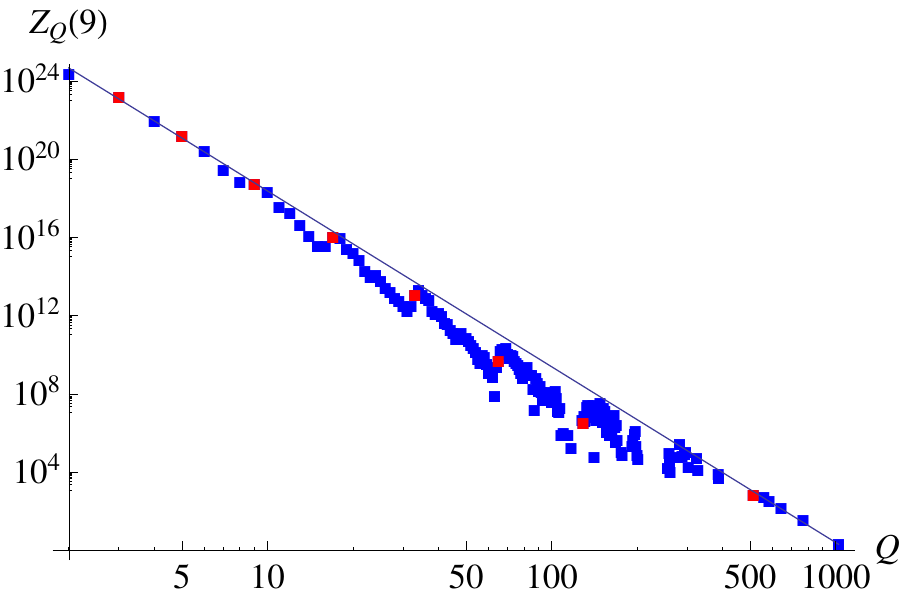}
\includegraphics[width=0.49\columnwidth]{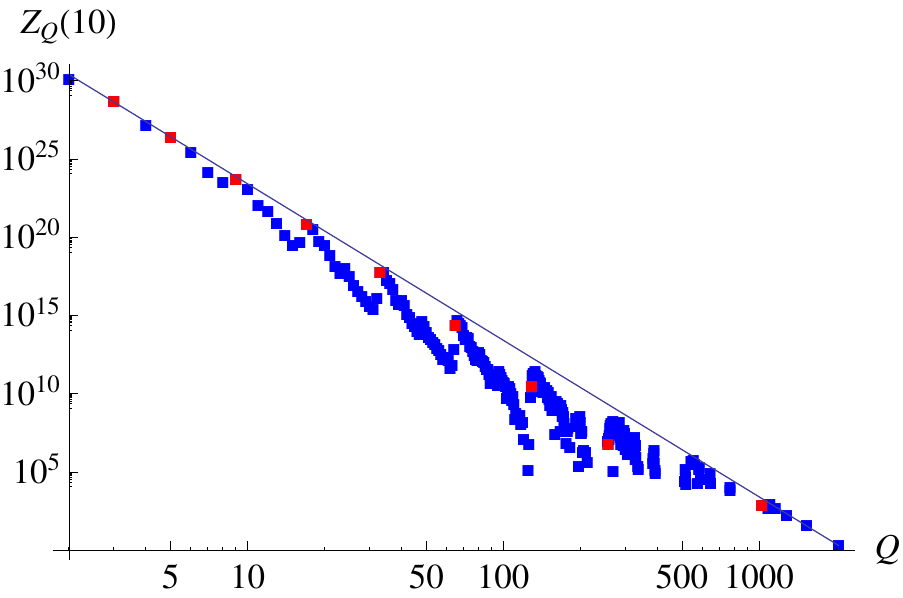}\\
\includegraphics[width=0.49\columnwidth]{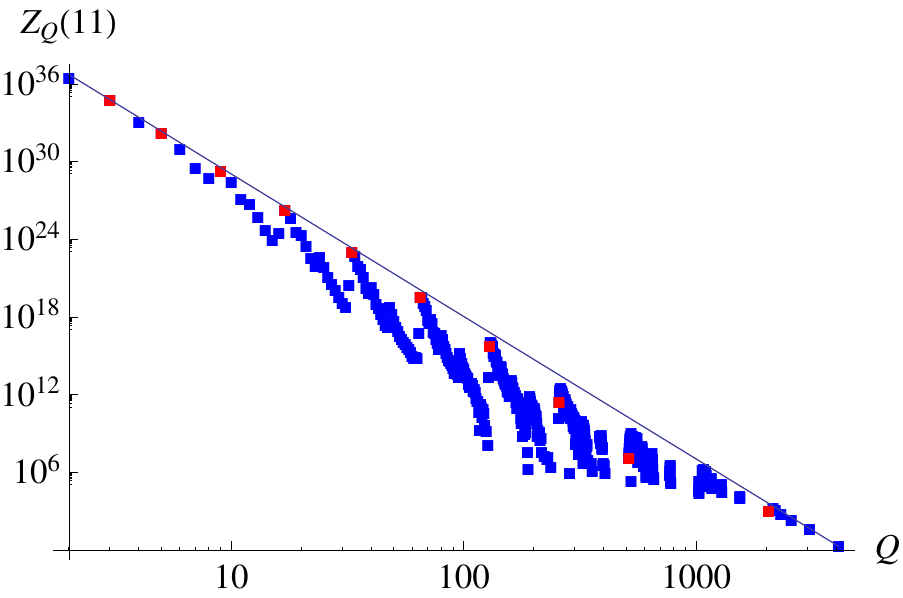}
\includegraphics[width=0.49\columnwidth]{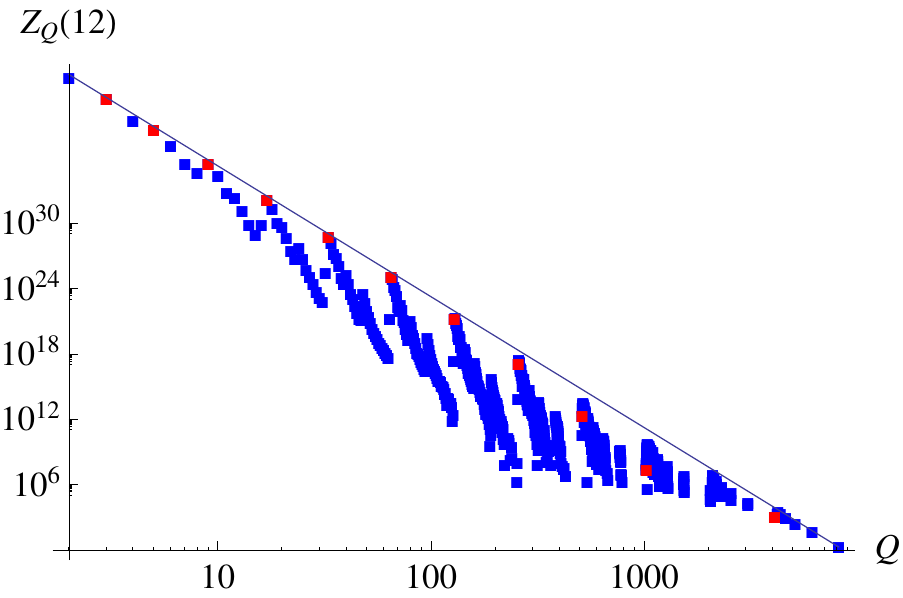}\\
\includegraphics[width=0.49\columnwidth]{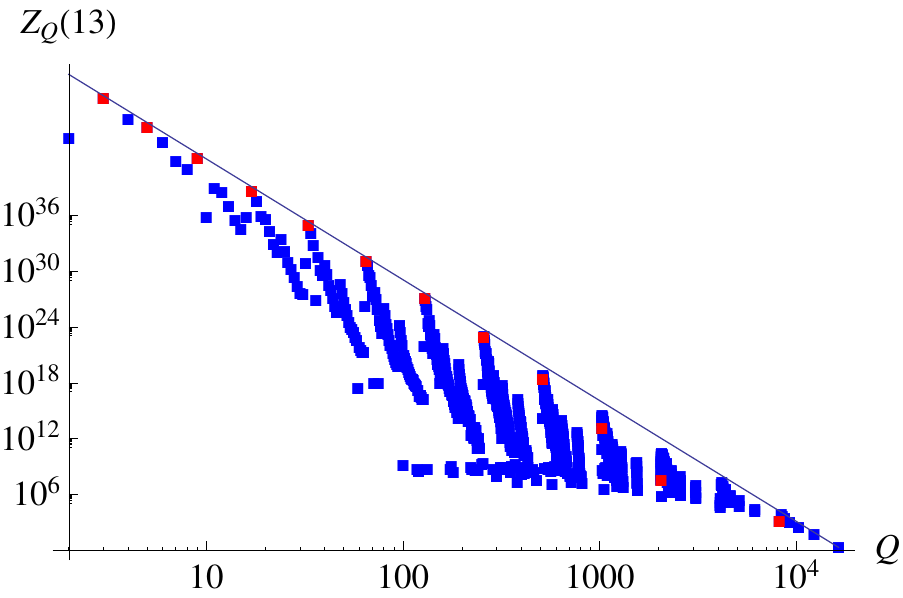}
\includegraphics[width=0.49\columnwidth]{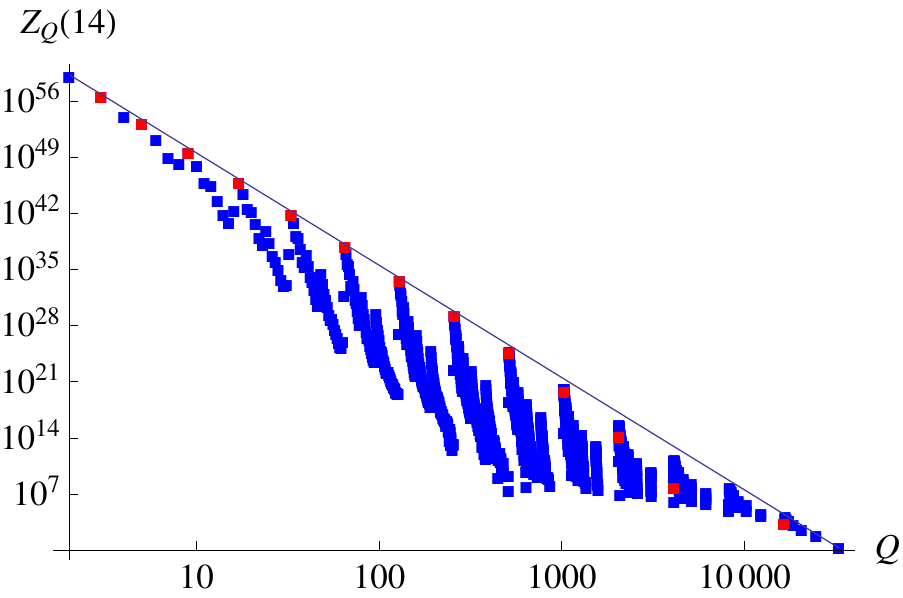}\\
\caption{The number of $\tilde{\sigma}_z$ textures which have $Q$ solutions, $Z_{Q}$, vs. $Q$. The continuous line is given by Eqs. (\ref{zapp}) and (\ref{eq:A_L}). The red dots mark the points at which $Q=2^{n}+1$.}
\label{zn}
\end{figure}

\begin{figure}
\includegraphics[width=\columnwidth]{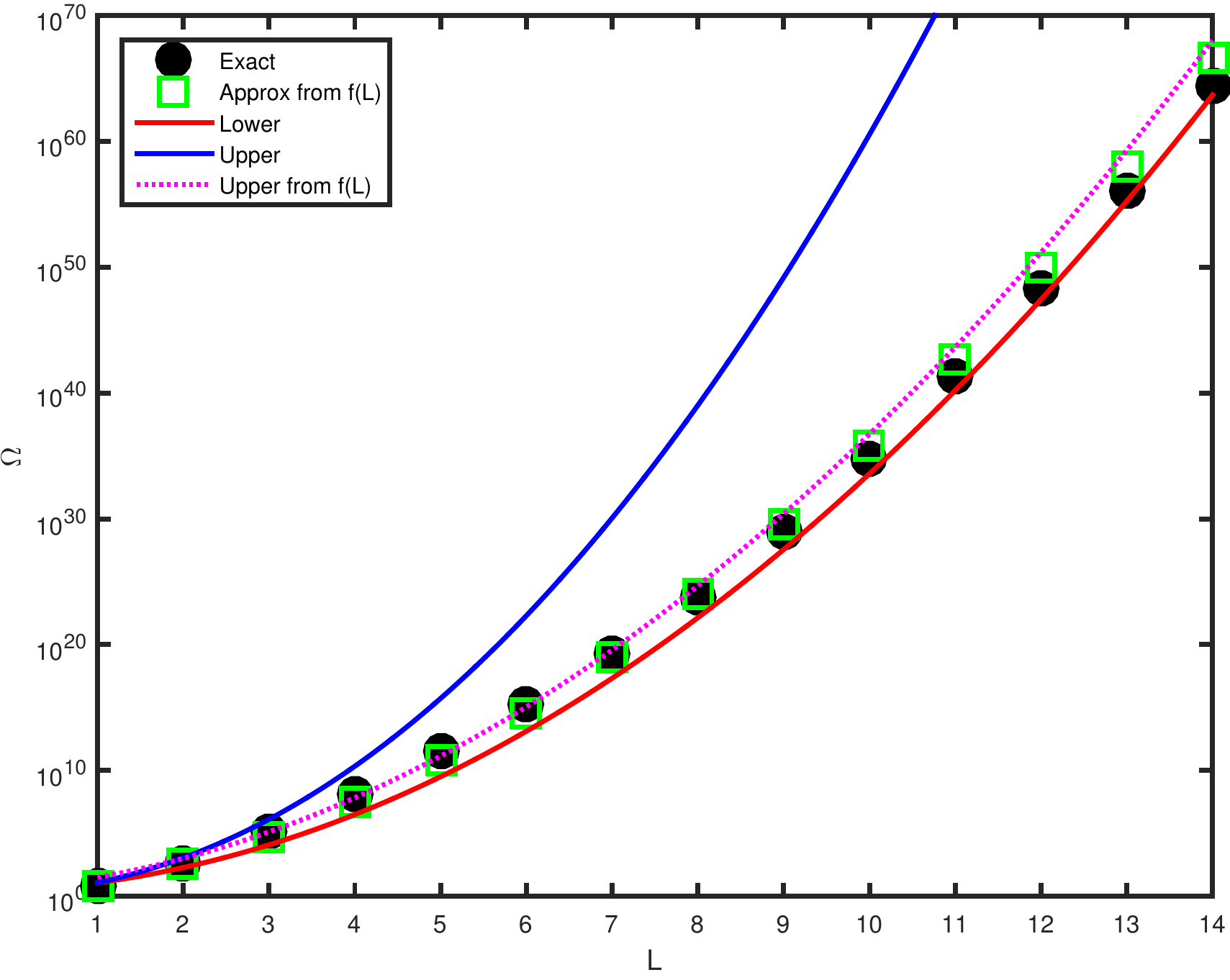}
\caption{Number of compatible configurations for a $L \times L \times L$ metacube. Black dots are the exact values. Solid red and blue lines indicate, respectively, the asymptotic $L \gg 1$ results for the lower (\ref{eq:lower_asymp}) and upper bounds (\ref{eq:upper_asymp2}). Green open squares are obtained by substituting in Eq.~(\ref{eq:omega_approx}) the exact values of $f(L)$, and the magenta dotted line is the approximate upper bound (\ref{eq:omega_approx_bound}).}
\label{fig:omega_vs_L}
\end{figure}

We numerically solved these recursion equations for $L\leq14$ and obtained the values of $Z_{Q}(L)$ shown in Fig. \ref{zn}. These were then used in Eq. (\ref{omdef}) to obtain the exact values of $\Omega(L)$ given in the main text and in Fig. \ref{fig:omega_vs_L}.

\subsection{Estimates}
\label{estimates}

\begin{figure}
\includegraphics[width=\columnwidth]{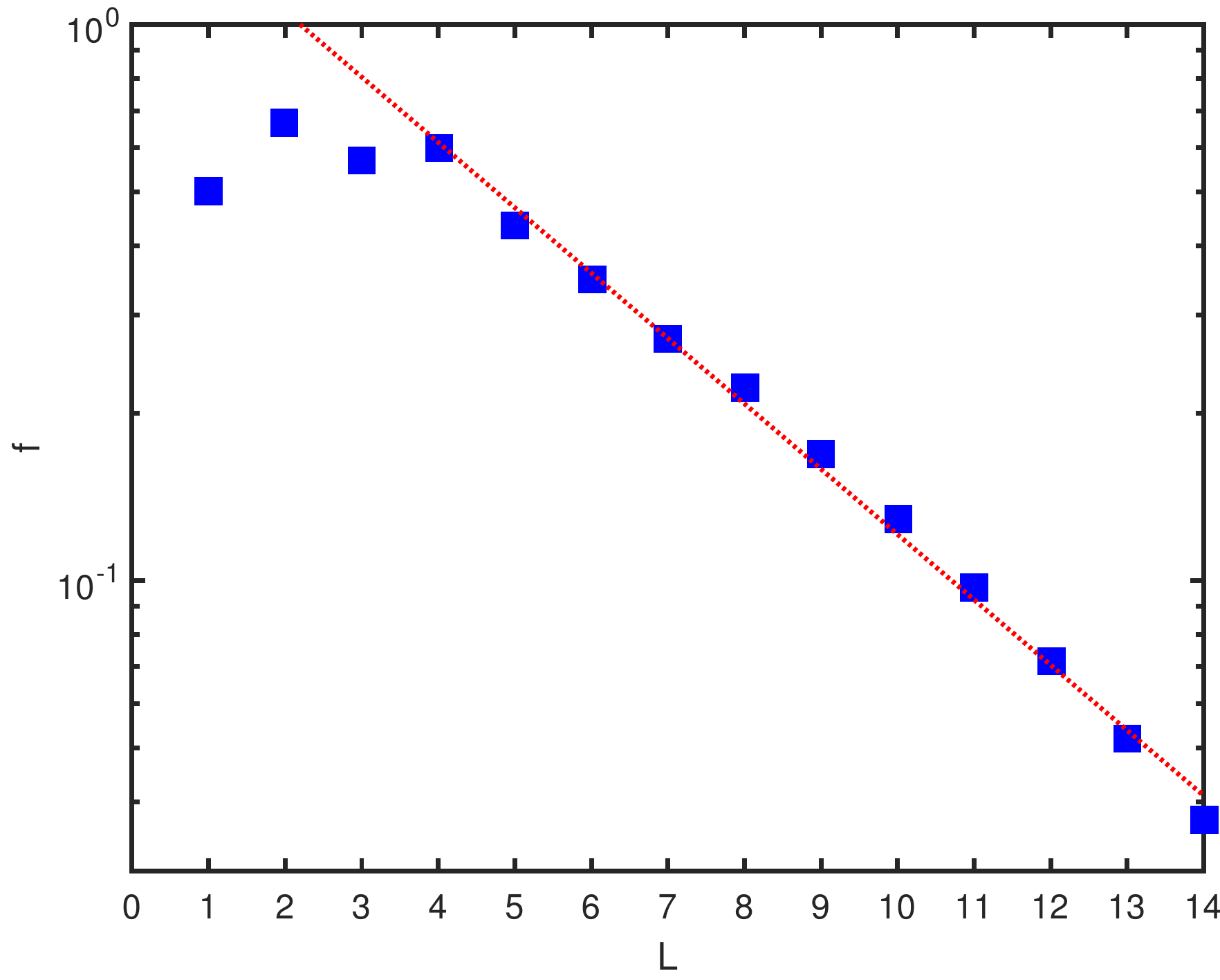}
\caption{The fraction $f(L)$ of $Q$ values for which there is at least one configuration with that number of solutions. The dashed line shows that for $L\geq4$, $f\approx1.8\times e^{-0.27L}$}
\label{f_vs_l}
\end{figure}

Note that although $2 \le Q \le 2^{L+1}-1$, for many values of $Q$ we have $Z_{Q}(L)=0$ and they are thus irrelevant. We denote by $f(L)$ the fraction of relevant $Q$ values for which $Z_{Q}(L)>0$. For $L \le 14$ we have exact values of $f(L)$, see Fig.~\ref{f_vs_l} and Table~\ref{f_vs_l_table}. Interestingly for $L\le8$ we numerically find that for the $Q$ values with $Z_{Q}(L)>0$ the following relation holds (see Fig.~\ref{zn}),
\begin{align}
Z_{Q}(L)\approx A(L)Q^{-L} . \label{zapp}
\end{align}
For $9\le L\le14$ we find that there are $L$ values of $Q$, located at approximately $Q=2^{n}+1$ $(1\leq n\leq L)$, for which $Z_{Q}(L)\approx A(L)Q^{-L}$, while for other values of $Q$ $Z_{Q}(L)\ll A(L)Q^{-L}$. These $L$ points contribute the most to $\Omega$. For example, for $L=14$ we find that including only the $14$ points for which $Q=2^{n}+1$ in Eq. (\ref{omdef}) yields a value which is about $0.36\times\Omega(14)$.

\begin{table}
\begin{tabular}{l|l}
\textbf{L}&\textbf{f(L)}\\\hline
1&1/2=0.5\\
2&4/6=0.67\\
3&8/14=0.57\\
4&18/30=0.6\\
5&27/63=0.44\\
6&44/126=0.35\\
7&69/254=0.27\\
8&113/510=0.22\\
9&172/1022=0.17\\
10&264/2046=0.13\\
11&399/4094=0.097\\
12&587/8190=0.072\\
13&852/16382=0.052\\
14&1213/32766=0.037
\end{tabular}
\caption{The fraction $f(L)$ of $Q$ values for which $Z_{Q}(L)>0$.}
\label{f_vs_l_table}
\end{table}

Assuming that (\ref{zapp}) holds, we can find the prefactor $A(L)$ analytically since there are two textures that have $Q_{max}=2^{L+1}-1$ solutions, 
\begin{align}
Z_{Q_{max}}(L) = A(L) \left(2^{L+1}-1\right)^{-L} = 2 ,
\end{align}
and thus
\begin{align}
A(L)=2\left(2^{L+1}-1\right)^{L} . \label{eq:A_L}
\end{align}

Therefore, we may approximate
\bea
\Omega(L) &\approx& \sum_{Q'} Z_Q(L) Q^L = A(L) \sum_{Q'} Q^{-L} Q^L \non\\ 
&=& f(L) A(L) \left( Q_{max} - 1 \right) \non\\
&=& f(L) \left(2^{L+1}-1\right)^{L} \cdot \left( 2^{L} - 1 \right) \cdot 4 , \label{eq:omega_approx}
\eea
where the sum over $Q'$ includes only the terms with $Z_Q(L)>0$. 

First, we verify that by substituting the exact values of $f(L)$ for $L \le 14$ in Eq.~(\ref{eq:omega_approx}) we indeed get a good approximation for $\Omega(L)$, see Fig.~\ref{fig:omega_vs_L}. Now, we use the fact that by definition $f(L) \le 1$ in order to obtain from the approximate result (\ref{eq:omega_approx}) the approximate upper bound
\bea
\Omega(L)\!\le\!\left(2^{L+1}\!-\!1\right)^{L}\! \cdot\! \left( 2^{L}\! -\! 1 \right) \!\cdot\! 4 \!\approx 2^{L^2+2L+2} , \label{eq:omega_approx_bound}
\eea
which is much tighter than (\ref{eq:upper_exact}), (\ref{eq:upper_asymp}) and (\ref{eq:upper_asymp2}), and rather close to the asymptotic lower bound (\ref{eq:lower_asymp}), see Fig.~\ref{fig:omega_vs_L}. Note that (\ref{eq:omega_approx_bound}) is only an approximate upper bound due to the approximation in (\ref{zapp}), from which it was derived. Combining Eqs. (\ref{eq:lower_asymp}) and (\ref{eq:omega_approx_bound}) we may write our tight bounds as:
\begin{align}
2^{L^2+L+log_2(3)} \leq \Omega \leq 2^{L^2+2L+2} .
\end{align}

\end{document}